\documentclass[a4paper]{aa}

\usepackage{graphicx,natbib}
\usepackage{txfonts}
\usepackage{color}
\usepackage[usenames,dvipsnames]{xcolor}

\newcommand{\bz}{$\langle B_\mathrm{z} \rangle$}
\newcommand{\hr}{HR\,5624}
\newcommand{\teff}{$T_{\rm eff}$}
\newcommand{\lgg}{$\log g$}
\newcommand{\vs}{$v_{\rm e}\sin i$}
\newcommand{\kms}{km\,s$^{-1}$}

\newcommand{\figps}[1]{\resizebox{\hsize}{!}{\rotatebox{0}{\includegraphics{#1}}}}

\newcommand{\fifps}[2]{\centering\resizebox{#1}{!}{\includegraphics{#2}}}
\newcommand{\firrps}[2]{\resizebox{#1}{!}{\rotatebox{270}{\includegraphics{#2}}}}

\newcommand{\beq}{\begin{equation}}
\newcommand{\eeq}{\end{equation}}

\begin{document}
\title{Magnetic field topology and chemical abundance distributions of the young, rapidly rotating chemically peculiar star HR\,5624\thanks{Based on observations collected at the European Southern Observatory, Chile (ESO programs 085.D-0296, 089.D-0383, 095.D-0194) and on observations obtained at the Canada-France-Hawaii Telescope (CFHT) which is operated by the National Research Council of Canada, the Institut National des Science de l'Univers of the Centre National de la Recherche Scientifique of France, and the University of Hawaii.}}

\author{O.~Kochukhov\inst{1}
   \and J.~Silvester\inst{1}
   \and J.D.~Bailey\inst{2}
   \and J.D.~Landstreet\inst{3,4}
   \and G.A.~Wade\inst{5}}

\institute{
Department of Physics and Astronomy, Uppsala University, Box 516, SE-75120 Uppsala, Sweden\\\email{oleg.kochukhov@physics.uu.se}
\and
Max Planck Insitut f\"ur Extraterrestrische Physik, Giessenbachstrasse 1, 85748 Garching, Germany
\and
Department of Physics and Astronomy, University of Western Ontario, London, Ontario, N6A 3K7, Canada
\and
Armagh Observatory, College Hill, Armagh, BT61 9DG, Northern Ireland, UK
\and
Department of Physics, Royal Military College of Canada, PO Box 17000, Stn Forces, Kingston, ON K7K 7B4, Canada
}

\date{Received 03 April 2017 / Accepted 12 May 2017}

\titlerunning{Magnetic field topology and chemical spots of HR\,5624}
\authorrunning{O.~Kochukhov et al.}

\abstract
{
The young, rapidly rotating Bp star \hr\ (HD\,133880) shows an unusually strong non-sinusoidal variability of its longitudinal magnetic field. This behaviour was previously interpreted as the signature of an exceptionally strong, quadrupole-dominated surface magnetic field geometry.
}
{
We studied the magnetic field structure and chemical abundance distributions of \hr\ with the aim to verify the unusual quadrupolar nature of its magnetic field and to investigate correlations between the field topology and chemical spots.
}
{
We analysed high resolution, time series Stokes parameter spectra of \hr\ with the help of a magnetic Doppler imaging inversion code based on detailed polarised radiative transfer modelling of the line profiles.
}
{
We refined the stellar parameters, revised the rotational period, and obtained new longitudinal magnetic field measurements. Our magnetic Doppler inversions reveal that the field structure of \hr\ is considerably simpler and the field strength is much lower than proposed by previous studies. We find a maximum local field strength of 12~kG and a mean field strength of 4~kG, which is about a factor of three weaker than predicted by quadrupolar field models. Our model implies that overall large-scale field topology of \hr\ is better described as a distorted, asymmetric dipole rather than an axisymmetric quadrupole. The chemical abundance maps of Mg, Si, Ti, Cr, Fe, and Nd obtained in our study are characterised by large-scale, high-contrast abundance patterns. These structures correlate weakly with the magnetic field geometry and, in particular, show no distinct element concentrations in the horizontal field regions predicted by theoretical atomic diffusion calculations.
}
{
We conclude that the surface magnetic field topology of \hr\ is not as unusual as previously proposed. Considering these results together with other recent magnetic mapping analyses of early-type stars suggests that predominantly quadrupolar magnetic field topologies, invoked to be present in a significant number of stars, probably do not exist in real stars. This finding agrees with an outcome of the MHD simulations of fossil field evolution in stably stratified stellar interiors.
}


\maketitle

\section{Introduction}
\label{intro}

A small fraction of the upper main sequence stars exhibit strong, globally-organised, stable surface magnetic fields \citep{donati:2009}. These fields, typically ranging between a few hundred G to several tens of kG in strength, are believed to be acquired by the stars at an earlier evolutionary stage rather than powered by a contemporary dynamo and are thus known as ``fossil'' fields. The phenomenon of surface fossil magnetism affects about 10 per cent of all main-sequence stars in the interval between early O and early F spectral types \citep{wolff:1968,donati:2009,wade:2012b,fossati:2015}. The exact mechanisms responsible for the generation of these fields and explaining their relatively low incidence are yet to be conclusively established. Plausible hypotheses include an amplification of the magnetic flux during the gravitational collapse of protostellar clouds \citep{moss:2003}, convective dynamos operating at the pre-main-sequence evolutionary stage \citep{alecian:2014}, or early stellar merger events \citep[][and references therein]{schneider:2016}. The presence of a strong magnetic field at the surfaces of early-type stars  facilitates formation of inhomogeneities in the form of chemical abundance spots in B--F-type stars (Ap/Bp stars) or corotating circumstellar clouds in O and early B-type stars. These structures, in turn, giving rise to conspicuous periodic photometric and spectroscopic variability.

Since the earliest studies of fossil magnetic fields \citep{stibbs:1950,deutsch:1958} it was evident that the majority of Ap/Bp stars show a single-wave, roughly sinusoidal periodic variation of the mean line-of-sight (longitudinal) magnetic field \bz. This observation was interpreted in the framework of the phenomenological oblique rotator model \citep{stibbs:1950}, which describes the stellar surface magnetic field topology as a frozen-in dipole, inclined with respect to the rotational axis of the star. This model is still widely used for statistical studies of large stellar samples \citep[e.g.][]{auriere:2007,hubrig:2007b} and as a starting guess for in-depth analyses of the field structure of individual stars. However, modelling \bz\ curves simultaneously with other integral magnetic observables (mean field modulus, crossover, mean quadratic field, broadband linear polarisation; see the comprehensive review by \citealt{mathys:2002}) required higher order multipolar field parameterisations in which quadrupolar terms often provided a significant, if not dominating, contribution \citep{landstreet:2000,bagnulo:2000,bagnulo:2001,bagnulo:2002}. Moreover, several early-type magnetic stars, including HD\,37776 \citep{thompson:1985}, \hr\ \citep{landstreet:1990}, $\tau$~Sco \citep{donati:2006b} and to a lesser extent HD\,32633 \citep{silvester:2012}, HD\,126515 \citep{mathys:2017} and HD\,175362 \citep{mathys:1997a}, show distinctly non-sinusoidal, double-wave \bz\ phase curves. In the literature, these observations are commonly considered as evidence of a field topology dominated by an axisymmetric quadrupolar component.

More recently, direct modelling of high resolution Stokes profile time series observations with the help of magnetic Doppler imaging \citep[MDI,][]{piskunov:2002a,kochukhov:2016} provided a new insight into fossil magnetic field geometries of early-type stars. Generally, the field topologies were found to be more diverse and occasionally much more complex than thought before. Inversions of high resolution Stokes parameter spectra, in particular those incorporating linear polarisation data, suggested the presence of small-scale magnetic concentrations as well as toroidal fields superimposed onto a global, dipole-like magnetic topology \citep{kochukhov:2004d,kochukhov:2010,silvester:2014,rusomarov:2016}. On the other hand, MDI of $\tau$~Sco and HD\,37776, suspected to host dominant quadrupolar fields according to their non-sinusoidal \bz\ variation, revealed highly structured, non-axisymmetric but weaker field geometries \citep{donati:2006b,kochukhov:2011a,kochukhov:2016a} than those inferred from the traditional low-order multipolar modelling of the integral magnetic observables. Thus, the question of the reality of strong, quadrupole-dominated fields in early-type stars is still unanswered. Do these field models provide a faithful representation of the actual surface field structures occurring in a small number of stars, or are they merely artefacts of parametric modelling limited to fitting the integral magnetic observables? The goal of this paper is to address this problem by investigating high resolution time series Stokes parameter observations of \hr, which belongs to a group of rare Ap/Bp stars (along with HD\,32633, HD\,37776, HD\,126515, HD\,175362, and $\tau$~Sco) with anharmonic longitudinal field curves and hence presumably very unusual underlying magnetic field structures.

The late B-type star \hr\ (HD\,133880, HIP\,74066, HR\,Lup) is a rapidly rotating, He-weak, Si-strong object. Despite its brightness ($V=5.8$), it was seldom analysed in detail due to its poor visibility from Northern hemisphere observatories. This star is a member of the Upper Cen-Lup association and thus has a well-established age of about 16 Myr \citep{landstreet:2007}. A strong magnetic field of \hr\ was detected by \citet{borra:1975} and further investigated by \citet{landstreet:1990} based on H$\beta$ longitudinal magnetic field measurements. The non-sinusoidal character of the \bz\ phase variation led \citet{landstreet:1990} to infer that the field topology is dominated by an axisymmetric quadrupole, which is at least 3 times stronger than the star's dipolar component. A detailed spectroscopic study of \hr\ was presented by \citet{bailey:2012}, who studied line profile variations, derived parameterised chemical abundance distributions and refined the magnetic field topology model based on longitudinal field and Zeeman broadening measurements. \citet{bailey:2012} concluded that \hr\ has an axisymmetric field structure described by a superposition of a 9.6~kG dipole and 23.2~kG quadrupole components, with the maximum local field strength exceeding 30~kG and the mean field modulus reaching 20~kG. This exceptionally strong and unusually complex magnetic field must produce high-amplitude circular and linear polarisation signatures in spectral lines, making this star an ideal target for an MDI analysis.

In this paper we present a new investigation of the surface magnetic field structure and chemical spot distributions of \hr\ using a superb spectropolarimetric data set collected for this star at the European Southern Observatory. The rest of the paper is organised as follows. Section~\ref{obs} describes acquisition and reduction of the new spectropolarimetric observations of \hr. Section~\ref{analysis} discusses analysis methodology and presents main results, including a revision of the stellar parameters (Sect.~\ref{atmos}), analysis of the mean polarisation profiles (Sect.~\ref{lsd}) and mean longitudinal magnetic field (Sect.~\ref{bz}), qualitative diagnostic with the help of cumulative Stokes $V$ profiles (Sect.~\ref{csv}), and, finally, tomographic reconstruction of the magnetic and chemical abundance maps (Sect.~\ref{mdi}). The results of our investigation are discussed in the context of theoretical magnetic field and atomic diffusion studies in Sect.~\ref{disc} and summarised in Sect.~\ref{conc}.

\section{Spectropolarimetric observations}
\label{obs}

\begin{table*}[!th]
\centering
\caption{Journal of HARPSpol observations of \hr. 
\label{tbl:obs}}
{\small
\begin{tabular}{ccccc|ccccc}
\hline\hline
UT date & HJD & Phase & S/N & Stokes &
UT date & HJD & Phase & S/N & Stokes \\
\hline
03-05-2010 & 2455319.7324 & 0.6996 & 226 & $IV$ & 12-05-2015 & 2457154.7623 & 0.9414 & 293 & $IV$ \\
06-08-2012 & 2456145.5016 & 0.7650 & 103 & $IV$ & 12-05-2015 & 2457154.7735 & 0.9542 & 297 & $IV$ \\
10-05-2015 & 2457152.5257 & 0.3925 & 179 & $IV$ & 12-05-2015 & 2457154.7847 & 0.9669 & 292 & $IV$ \\
10-05-2015 & 2457152.5369 & 0.4053 & 117 & $IV$ & 12-05-2015 & 2457154.7965 & 0.9804 & 295 & $IQ$ \\
10-05-2015 & 2457152.5487 & 0.4188 & 110 & $IQ$ & 12-05-2015 & 2457154.8077 & 0.9931 & 283 & $IQ$ \\
10-05-2015 & 2457152.5599 & 0.4315 & 165 & $IQ$ & 12-05-2015 & 2457154.8189 & 0.0059 & 283 & $IU$ \\
10-05-2015 & 2457152.5711 & 0.4443 & 150 & $IU$ & 12-05-2015 & 2457154.8300 & 0.0186 & 275 & $IU$ \\
10-05-2015 & 2457152.5823 & 0.4570 & 132 & $IU$ & 12-05-2015 & 2457154.8418 & 0.0320 & 251 & $IV$ \\
10-05-2015 & 2457152.5951 & 0.4716 & 108 & $IV$ & 12-05-2015 & 2457154.8530 & 0.0448 & 264 & $IV$ \\
10-05-2015 & 2457152.6063 & 0.4843 & 117 & $IV$ & 12-05-2015 & 2457154.8642 & 0.0575 & 238 & $IV$ \\
10-05-2015 & 2457152.6184 & 0.4981 & 106 & $IQ$ & 12-05-2015 & 2457154.8754 & 0.0703 & 220 & $IV$ \\
10-05-2015 & 2457152.6295 & 0.5108 & 131 & $IQ$ & 12-05-2015 & 2457154.8866 & 0.0830 & 215 & $IV$ \\
10-05-2015 & 2457152.6408 & 0.5236 & 144 & $IU$ & 12-05-2015 & 2457154.8978 & 0.0958 & 205 & $IV$ \\
10-05-2015 & 2457152.6519 & 0.5364 & 144 & $IU$ & 13-05-2015 & 2457156.4955 & 0.9166 & 203 & $IV$ \\
10-05-2015 & 2457152.6643 & 0.5505 &  91 & $IV$ & 14-05-2015 & 2457156.5067 & 0.9293 & 207 & $IV$ \\
10-05-2015 & 2457152.6755 & 0.5632 &  52 & $IV$ & 14-05-2015 & 2457156.5190 & 0.9433 & 208 & $IQ$ \\
10-05-2015 & 2457152.6883 & 0.5778 &  65 & $IQ$ & 14-05-2015 & 2457156.5301 & 0.9561 & 212 & $IQ$ \\
10-05-2015 & 2457152.6995 & 0.5906 &  70 & $IQ$ & 14-05-2015 & 2457156.5474 & 0.9757 & 139 &  $I$ \\
10-05-2015 & 2457152.7107 & 0.6033 &  61 & $IU$ & 14-05-2015 & 2457156.5590 & 0.9889 & 221 & $IV$ \\
10-05-2015 & 2457152.7219 & 0.6161 &  31 & $IU$ & 14-05-2015 & 2457156.5702 & 0.0017 & 210 & $IV$ \\
10-05-2015 & 2457152.7336 & 0.6295 & 165 & $IV$ & 14-05-2015 & 2457156.5814 & 0.0144 & 204 & $IV$ \\
10-05-2015 & 2457152.7448 & 0.6422 & 213 & $IV$ & 14-05-2015 & 2457156.5925 & 0.0272 & 200 & $IV$ \\
10-05-2015 & 2457152.7567 & 0.6558 & 203 & $IQ$ & 14-05-2015 & 2457156.6052 & 0.0417 & 260 & $IU$ \\
10-05-2015 & 2457152.7679 & 0.6685 & 198 & $IQ$ & 14-05-2015 & 2457156.6164 & 0.0544 & 255 & $IU$ \\
10-05-2015 & 2457152.7791 & 0.6813 & 201 & $IU$ & 14-05-2015 & 2457156.6290 & 0.0688 & 251 & $IQ$ \\
10-05-2015 & 2457152.7903 & 0.6941 & 191 & $IU$ & 14-05-2015 & 2457156.6402 & 0.0815 & 258 & $IQ$ \\
10-05-2015 & 2457152.8022 & 0.7076 & 206 & $IV$ & 14-05-2015 & 2457156.6514 & 0.0943 & 250 & $IU$ \\
10-05-2015 & 2457152.8134 & 0.7203 & 209 & $IV$ & 14-05-2015 & 2457156.6626 & 0.1070 & 249 & $IU$ \\
10-05-2015 & 2457152.8265 & 0.7354 & 200 & $IQ$ & 14-05-2015 & 2457156.6747 & 0.1208 & 259 & $IV$ \\
10-05-2015 & 2457152.8377 & 0.7481 & 210 & $IQ$ & 14-05-2015 & 2457156.6859 & 0.1336 & 251 & $IV$ \\
10-05-2015 & 2457152.8489 & 0.7609 & 165 & $IU$ & 14-05-2015 & 2457156.6971 & 0.1463 & 280 & $IV$ \\
10-05-2015 & 2457152.8601 & 0.7736 & 184 & $IU$ & 14-05-2015 & 2457156.7083 & 0.1591 & 286 & $IV$ \\
10-05-2015 & 2457152.8719 & 0.7871 & 171 & $IV$ & 14-05-2015 & 2457156.7194 & 0.1718 & 291 & $IV$ \\
10-05-2015 & 2457152.8831 & 0.7998 & 164 & $IV$ & 14-05-2015 & 2457156.7306 & 0.1845 & 280 & $IV$ \\
12-05-2015 & 2457154.5078 & 0.6514 & 153 & $IV$ & 14-05-2015 & 2457156.7433 & 0.1990 & 256 & $IQ$ \\
12-05-2015 & 2457154.5190 & 0.6641 & 170 & $IV$ & 14-05-2015 & 2457156.7545 & 0.2117 & 282 & $IQ$ \\
12-05-2015 & 2457154.5302 & 0.6769 & 133 & $IV$ & 14-05-2015 & 2457156.7671 & 0.2261 & 259 & $IV$ \\
12-05-2015 & 2457154.5414 & 0.6896 &  85 & $IV$ & 14-05-2015 & 2457156.7783 & 0.2388 & 233 & $IV$ \\
12-05-2015 & 2457154.5534 & 0.7033 &  57 & $IQ$ & 14-05-2015 & 2457156.7894 & 0.2516 & 226 & $IV$ \\
12-05-2015 & 2457154.6550 & 0.8191 & 255 & $IQ$ & 14-05-2015 & 2457156.8006 & 0.2643 & 207 & $IV$ \\
12-05-2015 & 2457154.6662 & 0.8318 & 271 & $IQ$ & 14-05-2015 & 2457156.8118 & 0.2771 & 178 & $IV$ \\
12-05-2015 & 2457154.6774 & 0.8446 & 285 & $IU$ & 14-05-2015 & 2457156.8230 & 0.2898 & 184 & $IV$ \\
12-05-2015 & 2457154.6858 & 0.8542 & 201 &  $I$ & 14-05-2015 & 2457156.8363 & 0.3049 & 156 & $IU$ \\
12-05-2015 & 2457154.7058 & 0.8771 & 281 & $IU$ & 14-05-2015 & 2457156.8474 & 0.3177 & 117 & $IU$ \\
12-05-2015 & 2457154.7170 & 0.8898 & 296 & $IU$ & 14-05-2015 & 2457156.8608 & 0.3329 & 170 & $IV$ \\
12-05-2015 & 2457154.7288 & 0.9032 & 297 & $IV$ & 14-05-2015 & 2457156.8720 & 0.3457 & 129 & $IV$ \\
12-05-2015 & 2457154.7400 & 0.9159 & 288 & $IV$ & 14-05-2015 & 2457156.8832 & 0.3584 & 101 & $IV$ \\
12-05-2015 & 2457154.7511 & 0.9287 & 282 & $IV$ & 14-05-2015 & 2457156.8944 & 0.3711 &  56 & $IV$ \\
\hline
\end{tabular}
\tablefoot{The columns give the UT date at mid-observation, the corresponding Heliocentric Julian date, rotational phase for $P_{\rm rot}=0.877483$~d, the median signal-to-noise ratio per 0.8~\kms\ pixel in the 490--500~nm wavelength region, and the Stokes parameters obtained. Observations listed here were determined from two sub-exposures with the exception of the two Stokes $I$ spectra (phases 0.8542 and 0.9757), which were derived from single exposures.}
}
\end{table*}

The spectra of \hr\ were obtained with the HARPSpol polarimeter \citep{snik:2011,piskunov:2011} that feeds the HARPS spectrometer \citep{mayor:2003} at the ESO 3.6 m telescope. This spectropolarimetric instrument allows one to obtain four Stokes parameter spectra at a resolving power of $\lambda/\Delta\lambda \approx $\,110\,000 with a complete wavelength coverage of the 386--691~nm range apart from 8~nm gap centred at 529~nm. HARPSpol is equipped with two independent devices for circular and linear polarisation measurements. Each of them splits the incoming light into a pair of beams with orthogonal linear polarisation that are injected into fibers, dispersed by the echelle spectrograph and recorded simultaneously on a 2K$\times$4K CCD mosaic.

Each Stokes parameter observation with HARPSpol is comprised of a sequence of either two or four sub-exposures between which the retarder waveplates are rotated by 90\degr\ (for Stokes $V$ observations) or by 45\degr\ (for Stokes $QU$ observations). This has an effect of exchanging positions of the orthogonally polarised spectra on the detector and enables application of a spectropolarimetric demodulation scheme known as the ``ratio method'' \citep{donati:1997,bagnulo:2009}. The two-exposure sequence yields a minimum data set for this demodulation procedure while four exposures provide a redundancy necessary for deriving the diagnostic null spectrum. To produce the latter, the sub-exposures are combined in such a way as to cancel the stellar polarisation signal, allowing to assess spurious polarisation that might be present in the data.

The optimal extraction and wavelength calibration of the HARPSpol spectra of \hr\ was carried out with the {\sc REDUCE} code\footnote{\url{http://www.astro.uu.se/~piskunov/RESEARCH/REDUCE/}} by \citet{piskunov:2002}. A custom set of {\sc IDL} procedures was then used to perform continuum normalisation and spectropolarimetric demodulation. We refer the reader to the paper by \citet{rusomarov:2013} for an in-depth discussion of the HARPSpol four Stokes parameter observing procedures and detailed description of our data reduction pipeline.

The observations of \hr\ were performed on the nights of May 10, 12, and 14, 2015. Initially, we planned to secure a complete rotational phase coverage in all four Stokes parameters and therefore obtained on the first observing night 6 Stokes $V$, 5 Stokes $Q$, and 5 Stokes $U$ observations, each consisting of four 450~s sub-exposures. However, having examined these data, we found that \hr\ exhibits a significantly lower polarisation amplitude than predicted by the previous field geometry models of this star, rendering the signal-to-noise ratio (S/N) in the linear polarisation profiles insufficient for detailed modelling. Consequently, we modified the observing strategy for the remainder of the observing run, putting more emphasis on the Stokes $V$ observations. The final set of Stokes parameter spectra was derived from pairs of sub-exposures in order to improve the phase resolution. Analysis of the diagnostic null profiles obtained from four sub-exposure sequences showed no evidence of a spurious polarisation in any of the HARPSpol Stokes parameter observations.

The resulting HARPSpol data set analysed in this paper includes 52 Stokes $V$ spectra (50 observations from our 2015 observing run plus two archival circular polarisation observations obtained in 2010 and 2012) as well as 21 Stokes $Q$, 21 Stokes $U$, and 96 intensity (Stokes $I$) measurements. Table~\ref{tbl:obs} provides information on each of these observations, including mid-exposure UT date, heliocentric Julian date, rotational phase, and representative S/N. We note that two of the intensity spectra listed in this table correspond to interrupted polarimetric sequences and therefore do not have associated polarisation spectra.

In addition to the HARPSpol data, we have used 14 ESPaDOnS Stokes $V$ spectra of \hr\ obtained in 2010--2012. These data, retrieved from the CFHT Science Archive\footnote{\url{http://www.cadc-ccda.hia-iha.nrc-cnrc.gc.ca/en/cfht/}}, were reduced by the UPENA pipeline running the {\sc Libre ESpRIT} software (an updated version of the code described by \citealt{donati:1997}). The ESPaDOnS spectra cover the 370--1050~nm wavelength interval at a resolution of about 65\,000. The first three of these spectra were analysed by \citet{bailey:2012}. Considering that the HARPSpol data have a superior resolution and adequate phase coverage, we have used the ESPaDOnS spectra of \hr\ only for the purpose of deriving additional longitudinal field measurements (see Sect.~\ref{bz}).

\section{Analysis and results}
\label{analysis}

\subsection{Stellar parameters}
\label{atmos}

Several determinations of the atmospheric and fundamental parameters of \hr\ are available in the literature. The stellar effective temperature was determined by \citet[][$12000\pm400$~K]{kochukhov:2006}, \citet[][$12000\pm400$~K]{landstreet:2007}, and \citet[][$11930\pm210$~K]{netopil:2008} based on the Geneva and Str\"omgren photometric calibrations appropriate for magnetic chemically peculiar stars. On the other hand, \citet{bailey:2012} refrained from applying an Ap-star correction to their photometric temperature estimate and adopted $T_{\rm eff}=13000\pm600$~K.

Based on the comparison with stellar evolutionary models and Hipparcos parallax, \citet{kochukhov:2006} established $\log L/L_\odot=2.07\pm0.10$, $R/R_\odot=2.51\pm0.33$, and $M/M_\odot=3.10\pm0.12$. \citet{landstreet:2007} used the fact that \hr\ is a member of the Upper Centaurus Lupus association and therefore has a well-established age of $\log t = 7.2\pm0.1$~yrs. These authors obtained $\log L/L_\odot=2.10\pm0.10$, $R/R_\odot=2.60\pm0.38$, and $M/M_\odot=3.20\pm0.15$, consistent with a young star that has completed about 5 per cent of its main sequence lifetime.

\begin{figure}[!th]
\centering
\fifps{8.5cm}{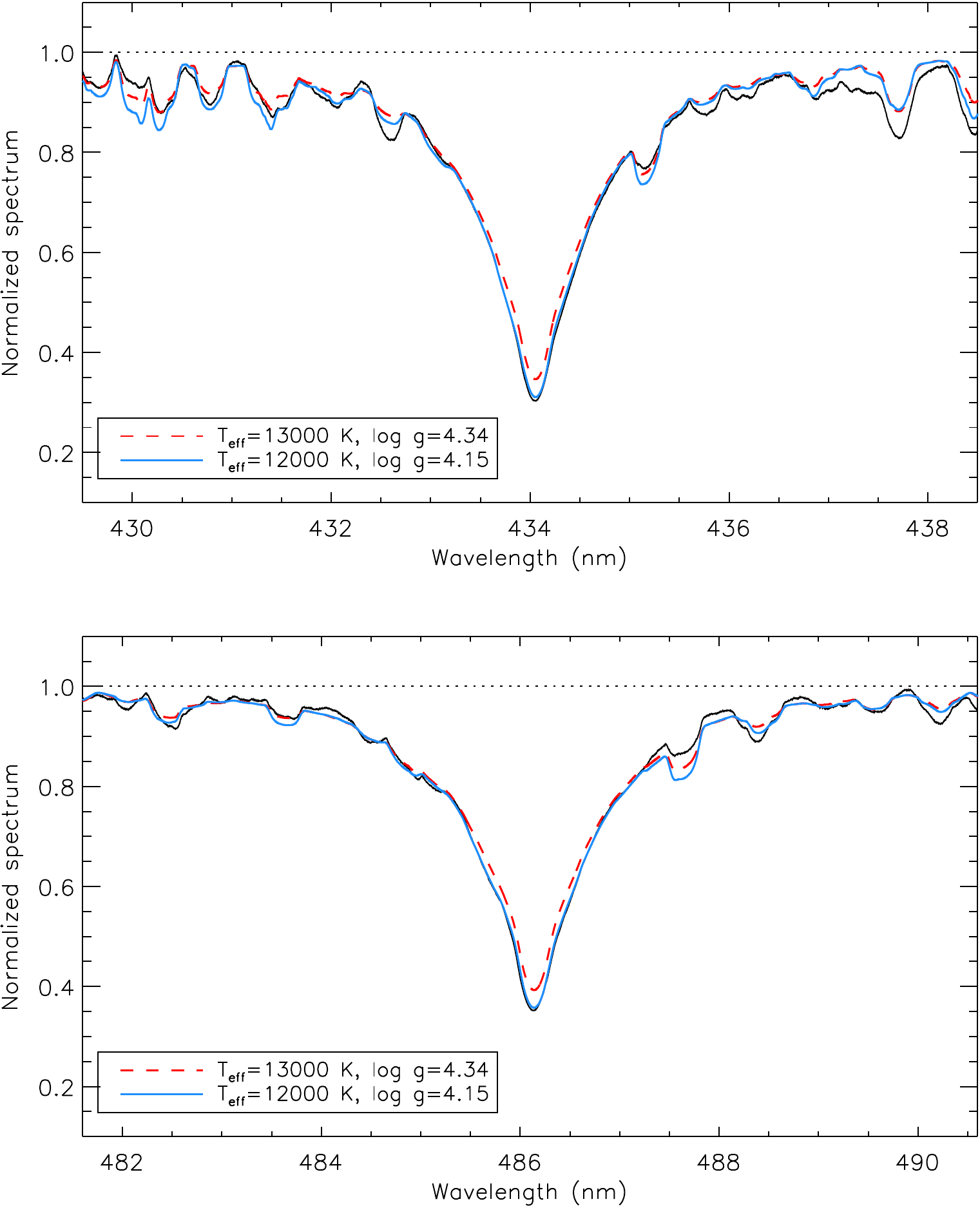}
\caption{Comparison of the mean observed hydrogen H$\gamma$ and H$\beta$ profiles (black histograms) with the synthetic spectra calculated for \teff\,=\,13000~K, \lgg\,=\,4.34 (dashed red line) and \teff\,=\,12000~K, \lgg\,=\,4.15 (solid blue line).}
\label{fig:hlines}
\end{figure}

In this paper we took advantage of the high resolution spectra around hydrogen Balmer lines to test different sets of atmospheric parameters. We have computed a grid of {\sc LLmodels} \citep{shulyak:2004} atmospheres using the average abundances reported by \citet{bailey:2012} and adopting a magnetic field of 5~kG to roughly account for the Zeeman desaturation of spectral lines. The hydrogen H$\beta$ and H$\gamma$ lines, computed for each model atmosphere in the grid using the {\sc Synmast} spectrum synthesis code \citep{kochukhov:2010a}, were compared with the average HARPSpol spectra to establish the best-fitting $T_{\rm eff}$ and $\log g$. As demonstrated by Fig.~\ref{fig:hlines}, $T_{\rm eff}=12000$~K and $\log g=4.15$ provides a better description of the observations compared to $T_{\rm eff}=13000$~K and $\log g=4.34$ used by \citet{bailey:2012}. Based on this comparison, we adopted the former set of atmospheric parameters for further analysis. We note that our $\log g$ established from fitting the hydrogen Balmer lines is in good agreement with the surface gravity of $\log g=4.11$--4.13 that can be inferred from the stellar evolution analyses by \citet{kochukhov:2006} and \citet{landstreet:2007}.


\subsection{Least-squares deconvolved profiles}
\label{lsd}

The Stokes $V$ profiles of strong metal lines in the spectrum of \hr\ exhibit distinct circular polarisation signatures. However, their relatively low S/N and often severe blending renders individual lines unsuitable for detailed modelling. No linear polarisation signatures could be detected in any of the individual spectral lines. Considering these circumstances, we applied the least-squares deconvolution (LSD) multi-line method \citep{donati:1997}, implemented as described by \citet{kochukhov:2010a}, in order to obtain high quality mean Stokes profiles for all metal lines and for subsets of a few selected chemical elements. The line parameters required for building an LSD mask were obtained from the VALD3 database \citep{ryabchikova:2015} using the $T_{\rm eff}=12000$~K, $\log g=4.15$ model atmosphere described above and the mean abundances from \citet{bailey:2012}. We then retained lines deeper than 5 per cent of the continuum (before any macroscopic broadening) and excluded all lines located in the wings of the hydrogen Balmer lines or in the regions of significant telluric absorption. Four sets of Stokes $IQUV$ LSD profiles were derived: i) using all 1916 available metal lines, ii) using 1127 Fe lines, iii) using 275 Cr lines, and iv) using 167 Si lines. The LSD line weights for the complete metal line mask were normalised using a mean wavelength $\lambda_0=523$~nm,  Land\'e factor $z_0=1.18$, and depth $d_0=0.18$. A common normalisation with $\lambda_0=518$~nm, $z_0=1.20$, $d_0=0.17$ was adopted for the Fe, Cr, and Si LSD profiles. As discussed by \citet{kochukhov:2010a}, the choice of LSD line weight normalisation is irrelevant as long as it is applied consistently in derivation of LSD profiles and in their interpretation. The same line lists and deconvolution procedures were applied to both HARPSpol and ESPaDOnS spectra; the LSD velocity step was chosen to be 1.6~\kms\ for HARPSpol and 2.6~\kms\ for the ESPaDOnS data, respectively. For consistency, the LSD analysis of the ESPaDOnS spectra was restricted to the HARPS wavelength range.

\begin{figure}[!th]
\centering
\figps{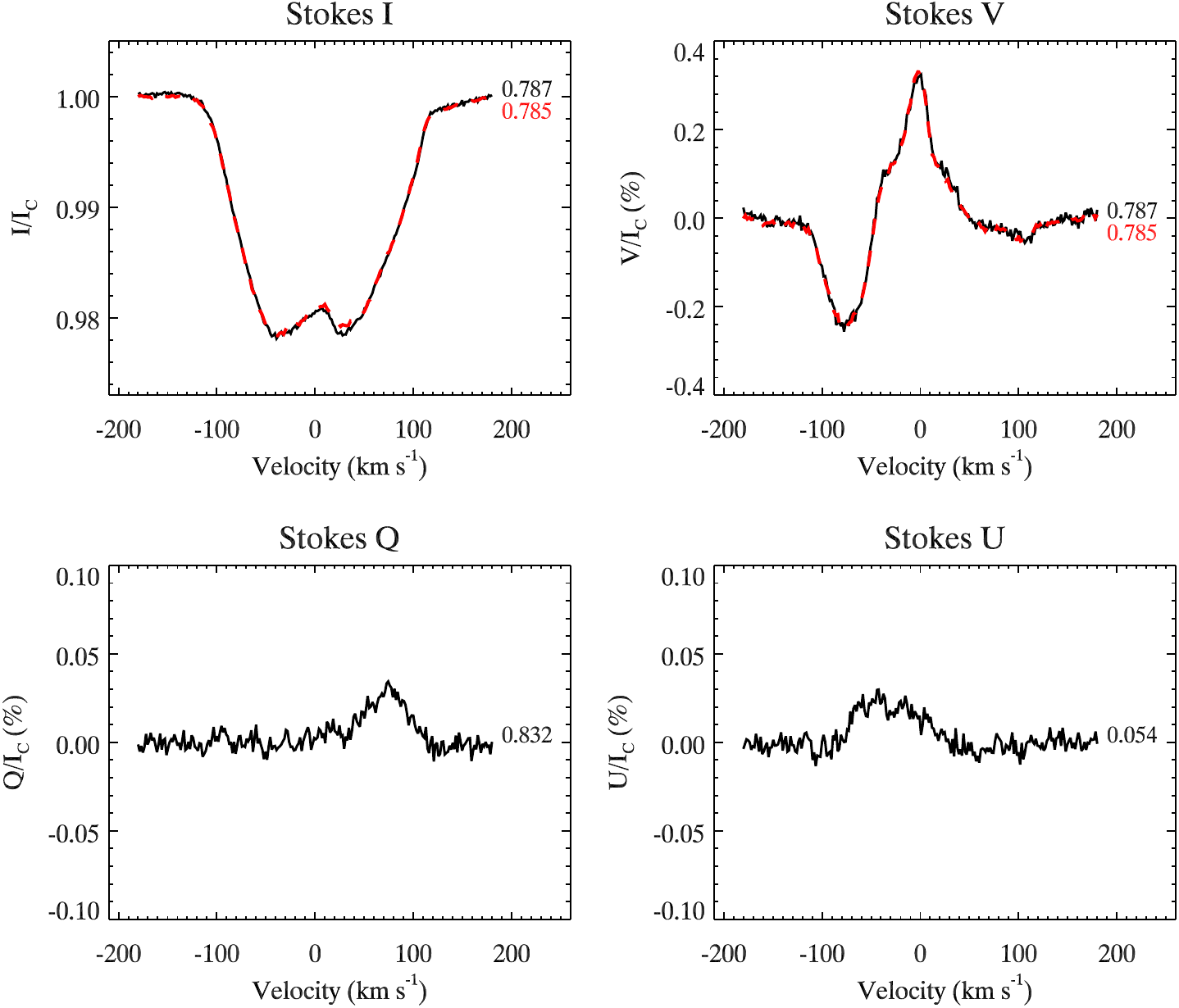}
\caption{Example of the four Stokes parameter LSD profiles derived for \hr\ using a complete metal line mask. The top panels compare the Stokes $I$ and $V$ HARPSpol profiles (solid lines) with the ESPaDOnS LSD profiles (dashed lines) obtained at close rotational phases. The bottom panels show examples of some of the strongest Stokes $Q$ and $U$ HARPSpol LSD profiles.}
\label{fig:lsd_sel}
\end{figure}

\begin{figure}[!th]
\centering
\figps{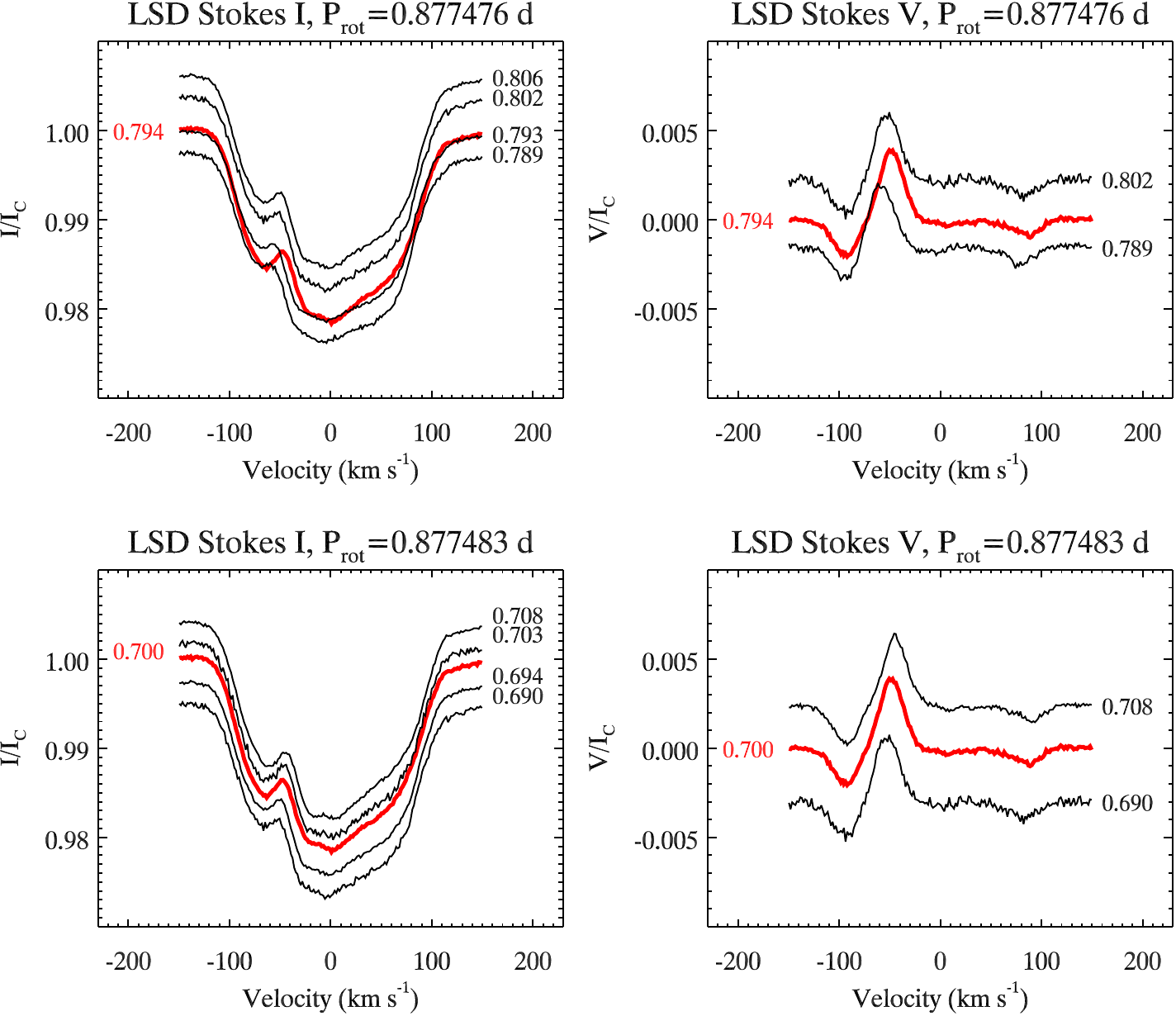}
\caption{Relative phasing of the LSD Stokes $I$ (left) and $V$ (right) profiles corresponding to the archival HARPSpol observation of \hr\  obtained in 2010 (thick, light/red curves) and observations collected in 2015 (thin, dark/black curves) for the rotational periods 0.877476 (top) and 0.877483~days (bottom). Profiles are shifted vertically according to the rotational phases, which are indicated to the right of the profiles for the 2015 data and to the left for the 2010 spectrum.}
\label{fig:phas}
\end{figure}

\begin{figure}[!th]
\centering
\fifps{7.5cm}{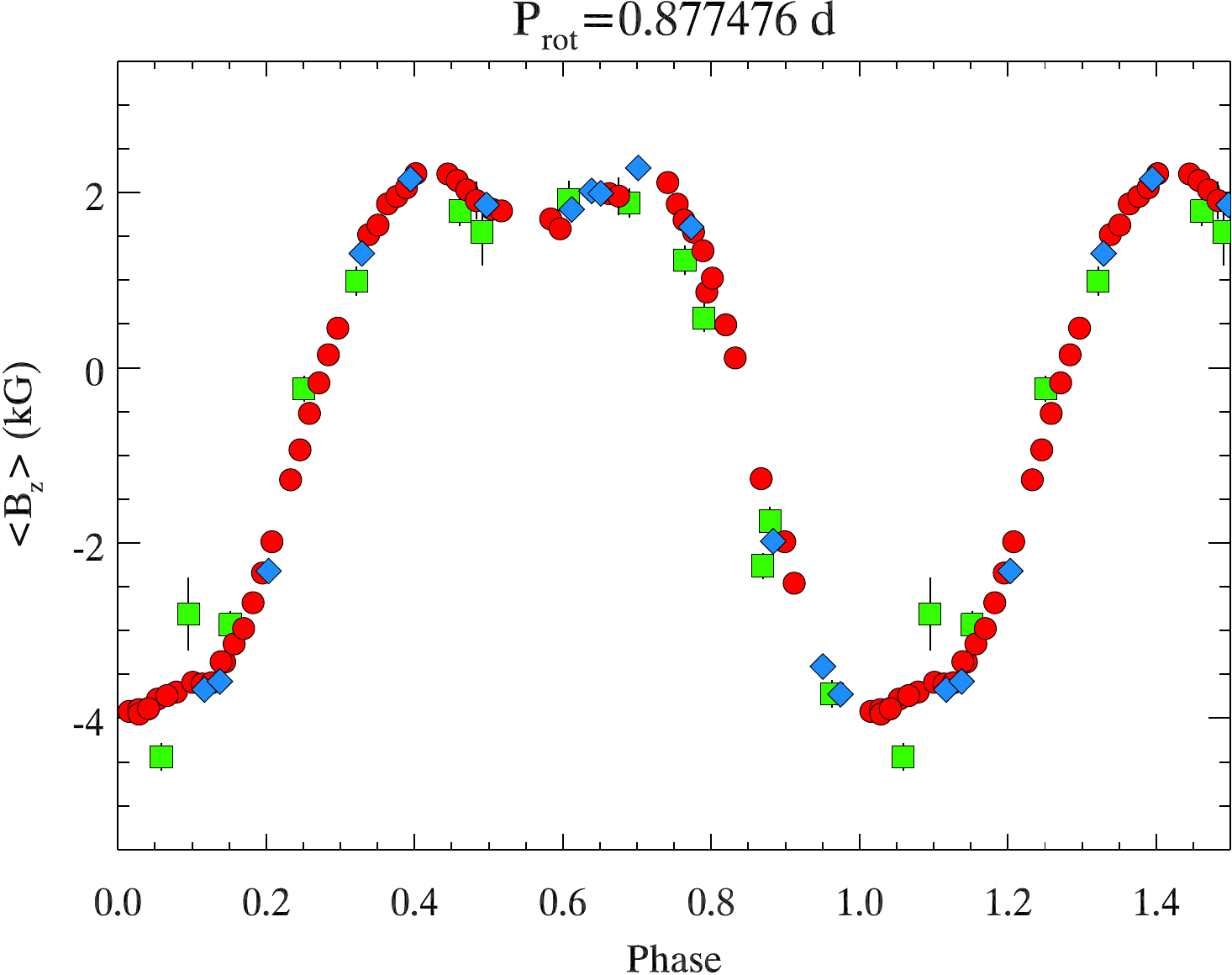}\\\vspace*{0.5cm}
\fifps{7.5cm}{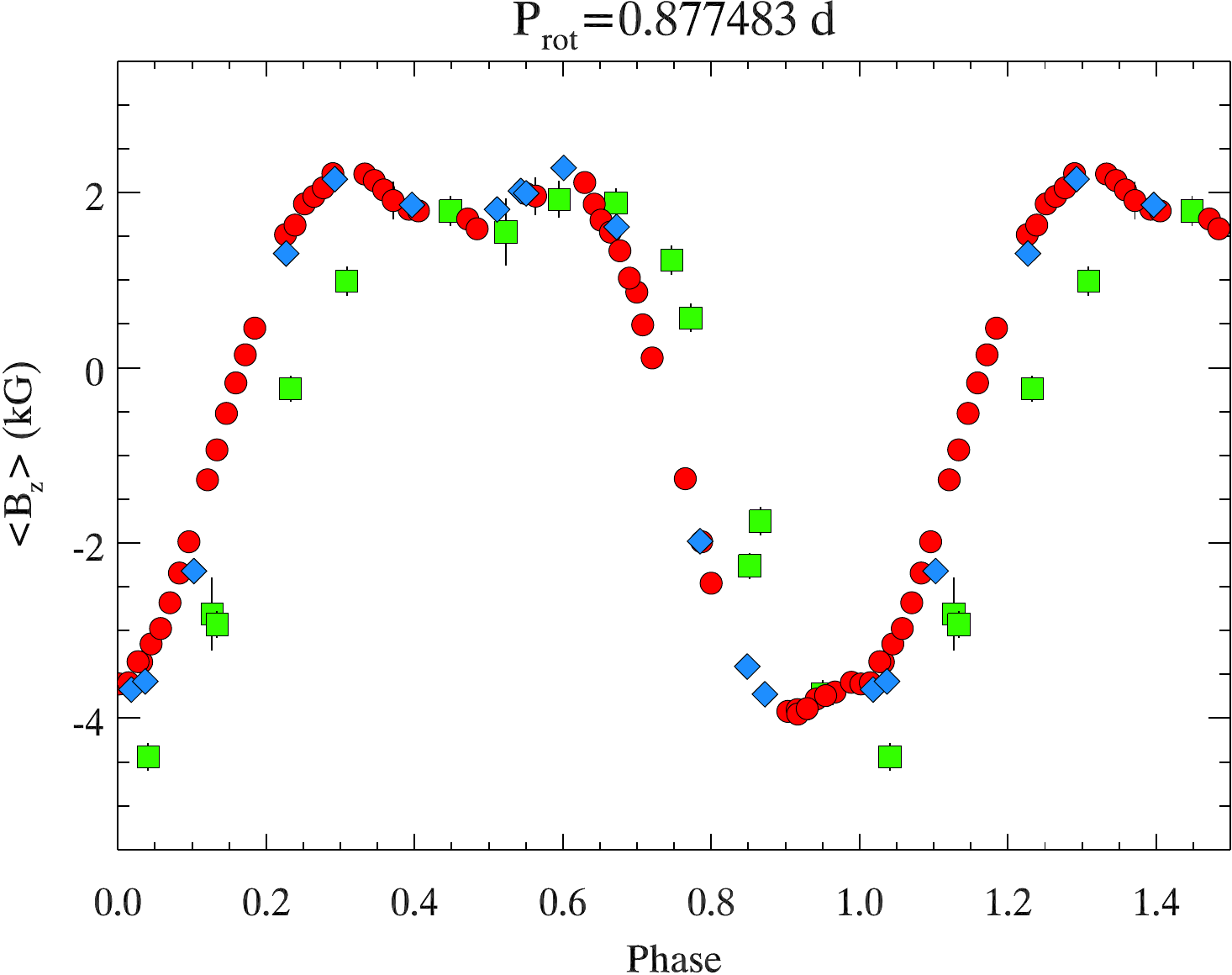}
\caption{Variation of the mean longitudinal magnetic field of \hr. The dark symbols show \bz\ measurements derived in this study from the HARPSpol (red circles) and ESPaDOnS (blue rhombs) observations. The green squares show the Balmer line photopolarimetric measurements by \citet{landstreet:1990}. The upper and lower panels illustrate phasing of the \bz\ data with the rotational periods 0.877476~d and 0.877483~d, respectively.}
\label{fig:bz}
\end{figure}

Figure~\ref{fig:lsd_sel} shows an example of the four Stokes parameter LSD profiles derived from the HARPSpol observations. For comparison, we also present in the upper panels of this figure the Stokes $I$ and $V$ LSD profiles corresponding to a nearby rotational phase observed by ESPaDOnS. The circular polarisation signal is detected with a high S/N, which is also the case for the Fe and Si LSD profiles. On the other hand, the Stokes $Q$ and $U$ signals are detected at a rather low S/N only in the LSD profiles corresponding to the complete metal line mask. The bottom panels of Fig.~\ref{fig:lsd_sel} display some of the strongest linear polarisation mean line signatures detected in the HARPSpol data.

A dense phase coverage secured by our HARPSpol observations, combined with plenty of sharp details evident in the Stokes $I$ and $V$ LSD profiles, enable an accurate determination of the stellar rotational period. Previous period determinations for \hr\ were summarised by \citet{bailey:2012}, who converged on the value of $0.877476\pm0.000009$~d. Considering phasing of the 2010 archival HARPSpol spectrum relative to our 2015 data set, we find a slight offset (Fig.~\ref{fig:phas}, upper panels). A revised period of 0.877483~d, which is only $0.8\sigma$ longer than the value preferred by \citet{bailey:2012}, yields a smooth variation of the Stokes $I$ and $V$ profiles with a minimum phase dispersion (Fig.~\ref{fig:phas}, lower panels). The same period also improves the phasing of the 2010--2012 ESPaDOnS spectra relative to the 2015 HARPSpol data. Therefore, we adopted this period and the reference Julian date 2445472.000 \citep{bailey:2012} corresponding to the minimum photometric brightness for the MDI modelling presented below.

\subsection{Mean longitudinal magnetic field}
\label{bz}

\begin{table*}[!th]
\centering
\caption{Mean longitudinal magnetic field of \hr. 
\label{tbl:bz}}
{\small
\begin{tabular}{crc|crc|crc}
\hline\hline
HJD & \bz\ (G)~~ & Inst &
HJD & \bz\ (G)~~ & Inst &
HJD & \bz\ (G)~~ & Inst \\
\hline
2455319.7324 & $  863\pm  46$ & H & 2457154.5190 & $ 1548\pm  59$ & H & 2457156.5067 & $-3890\pm  42$ & H \\
2456145.5016 & $-1266\pm  89$ & H & 2457154.5302 & $ 1335\pm  74$ & H & 2457156.5590 & $-3590\pm  41$ & H \\
2457152.8134 & $  112\pm  48$ & H & 2457154.5414 & $ 1023\pm 111$ & H & 2457156.5702 & $-3609\pm  43$ & H \\
2457152.8719 & $-1986\pm  54$ & H & 2457154.7288 & $-3922\pm  33$ & H & 2457156.5814 & $-3595\pm  45$ & H \\
2457152.8831 & $-2459\pm  56$ & H & 2457154.7400 & $-3902\pm  34$ & H & 2457156.5925 & $-3355\pm  46$ & H \\
2457152.5257 & $ 1809\pm  69$ & H & 2457154.7511 & $-3894\pm  34$ & H & 2457156.6747 & $-1279\pm  40$ & H \\
2457152.5369 & $ 1790\pm 105$ & H & 2457154.7623 & $-3782\pm  33$ & H & 2457156.6859 & $ -937\pm  41$ & H \\
2457152.5951 & $ 1698\pm 113$ & H & 2457154.7735 & $-3742\pm  33$ & H & 2457156.6971 & $ -521\pm  38$ & H \\
2457152.6063 & $ 1586\pm 104$ & H & 2457156.7083 & $ -173\pm  38$ & H & 2455411.7310 & $ 2018\pm  42$ & E \\
2457152.6643 & $ 1987\pm 127$ & H & 2457156.7194 & $  148\pm  37$ & H & 2455756.7940 & $-1980\pm  45$ & E \\
2457152.6755 & $ 1959\pm 217$ & H & 2457156.7306 & $  452\pm  38$ & H & 2455757.8761 & $-3673\pm  47$ & E \\
2457152.7336 & $ 2114\pm  64$ & H & 2457156.7671 & $ 1519\pm  42$ & H & 2455960.1644 & $ 1991\pm  47$ & E \\
2457152.7448 & $ 1867\pm  50$ & H & 2457156.7783 & $ 1628\pm  47$ & H & 2455961.0862 & $ 2281\pm  46$ & E \\
2457152.8022 & $  490\pm  48$ & H & 2457156.7894 & $ 1870\pm  49$ & H & 2455961.1488 & $ 1607\pm  47$ & E \\
2457154.7847 & $-3701\pm  34$ & H & 2457156.8006 & $ 1956\pm  54$ & H & 2455966.1717 & $ 1859\pm  84$ & E \\
2457154.8418 & $-3359\pm  39$ & H & 2457156.8118 & $ 2054\pm  63$ & H & 2455967.1499 & $ 1807\pm  38$ & E \\
2457154.8530 & $-3152\pm  38$ & H & 2457156.8230 & $ 2215\pm  62$ & H & 2455971.1211 & $-3580\pm  43$ & E \\
2457154.8642 & $-2976\pm  42$ & H & 2457156.8608 & $ 2211\pm  72$ & H & 2455971.1786 & $-2320\pm  52$ & E \\
2457154.8754 & $-2682\pm  45$ & H & 2457156.4955 & $-3953\pm  43$ & H & 2456108.7415 & $-3726\pm  39$ & E \\
2457154.8866 & $-2343\pm  46$ & H & 2457156.8720 & $ 2137\pm  97$ & H & 2456111.7426 & $ 2153\pm  40$ & E \\
2457154.8978 & $-1986\pm  48$ & H & 2457156.8832 & $ 2030\pm 125$ & H & 2456115.7405 & $-3410\pm  39$ & E \\
2457154.5078 & $ 1685\pm  66$ & H & 2457156.8944 & $ 1908\pm 219$ & H & 2456117.8275 & $ 1304\pm  48$ & E \\
\hline
\end{tabular}
\tablefoot{The columns give the Julian date, the mean longitudinal magnetic field measurement and the corresponding error bar. Measurements derived from HARPSpol (ESPaDOnS) spectra are indicated with ``H'' (``E'').}
}
\end{table*}

The mean longitudinal magnetic field \bz\ measures the disk-averaged line-of-sight magnetic field component, weighted by the local continuum brightness and line strength. This observable is commonly used for characterising magnetic fields of Ap/Bp stars. The non-sinusoidal variation of \bz\ in \hr\ was considered as the basic primary evidence of a quadrupole-dominated magnetic field topology in this star \citep{landstreet:1990}.

We have inferred the mean longitudinal magnetic field of \hr\ from the Stokes $I$ and $V$ LSD profiles corresponding to the complete metal line mask. These \bz\ measurements were derived by calculating the first moment of the Stokes $V$ profile, normalised by the equivalent width of the corresponding Stokes $I$ profile and scaled by the appropriate mean wavelength and Land\'e factor \citep{kochukhov:2010a}. The numerical integration was carried out in the velocity range between $-120$ and $+125$~\kms\ and was applied consistently to both HARPSpol and ESPaDOnS data. The resulting \bz\ values, summarised in Table~\ref{tbl:bz}, indicate variation between about $-3.9$ and $+2.3$~kG. The median error bar of our measurements is 47~G.

The phase variation of the HARPSpol and ESPaDOnS \bz\ measurements agree very well. These data are also compatible, despite a difference in measurement methodology, with the hydrogen line photopolarimetric longitudinal field measurements collected by \citet{borra:1975} and \citet{landstreet:1990} provided that $P_{\rm rot}=0.877476$~d is used. However, for our preferred $P_{\rm rot}=0.877483$~d there is a 0.07 phase offset between the photopolarimetric data from 1974--1988 and the new spectropolarimetric measurements obtained in 2010--2015. This offset cannot be removed by postulating a simple linear period increase corresponding to the canonical rotational braking \citep{ud-doula:2009}. Instead, a step-like or continuous period \textit{decrease}, comparable to the one experienced by CU~Vir \citep{mikulasek:2011}, might be required to reconcile \bz\ and line profile variation. 

\subsection{Cumulative Stokes $V$ profiles}
\label{csv}

\begin{figure}[!th]
\includegraphics[height=8.3cm]{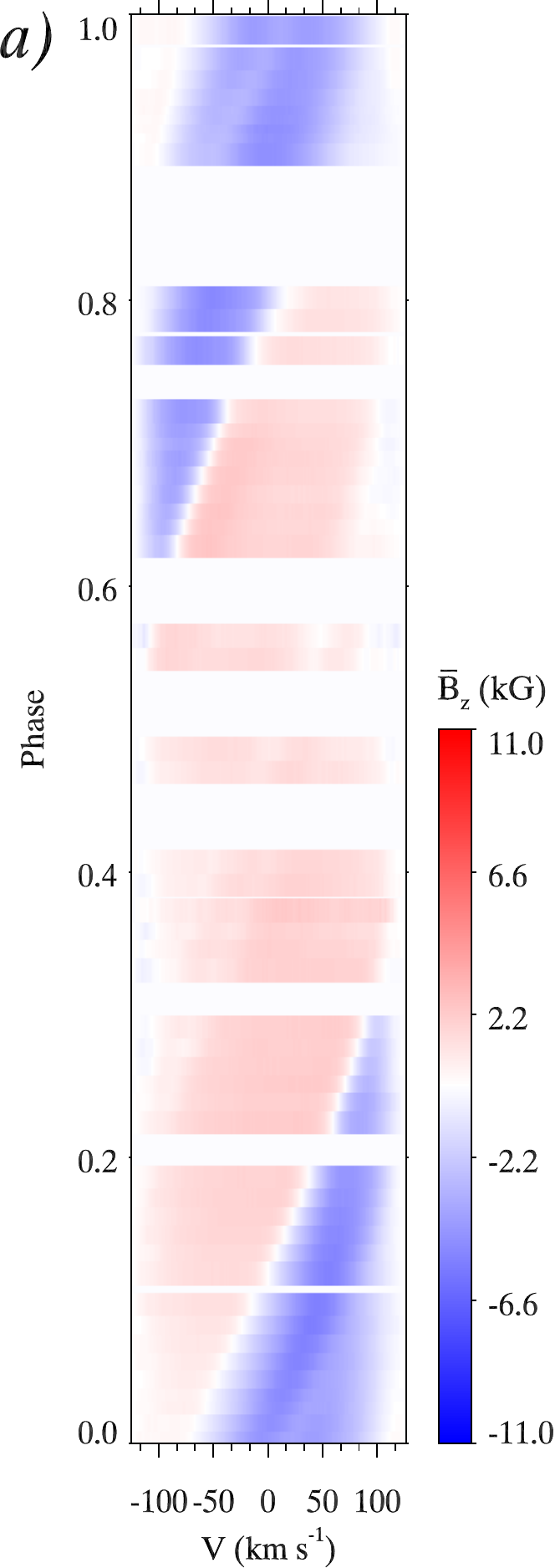}
\includegraphics[height=8.3cm]{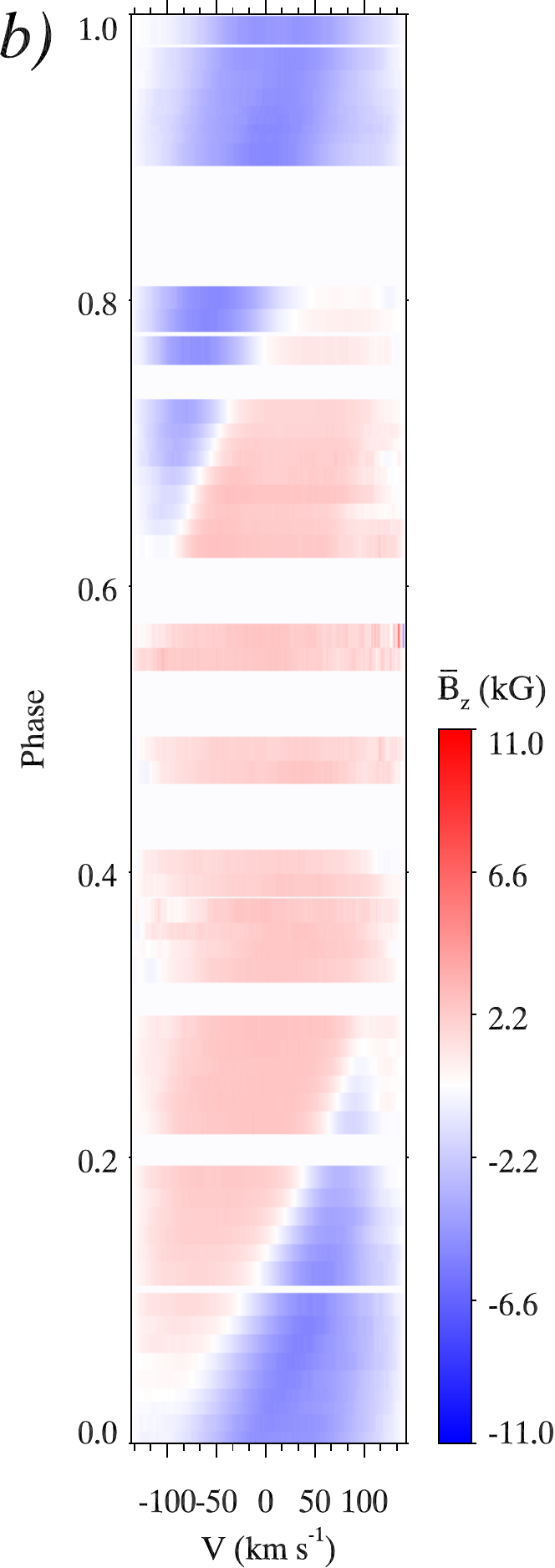}
\includegraphics[height=8.3cm]{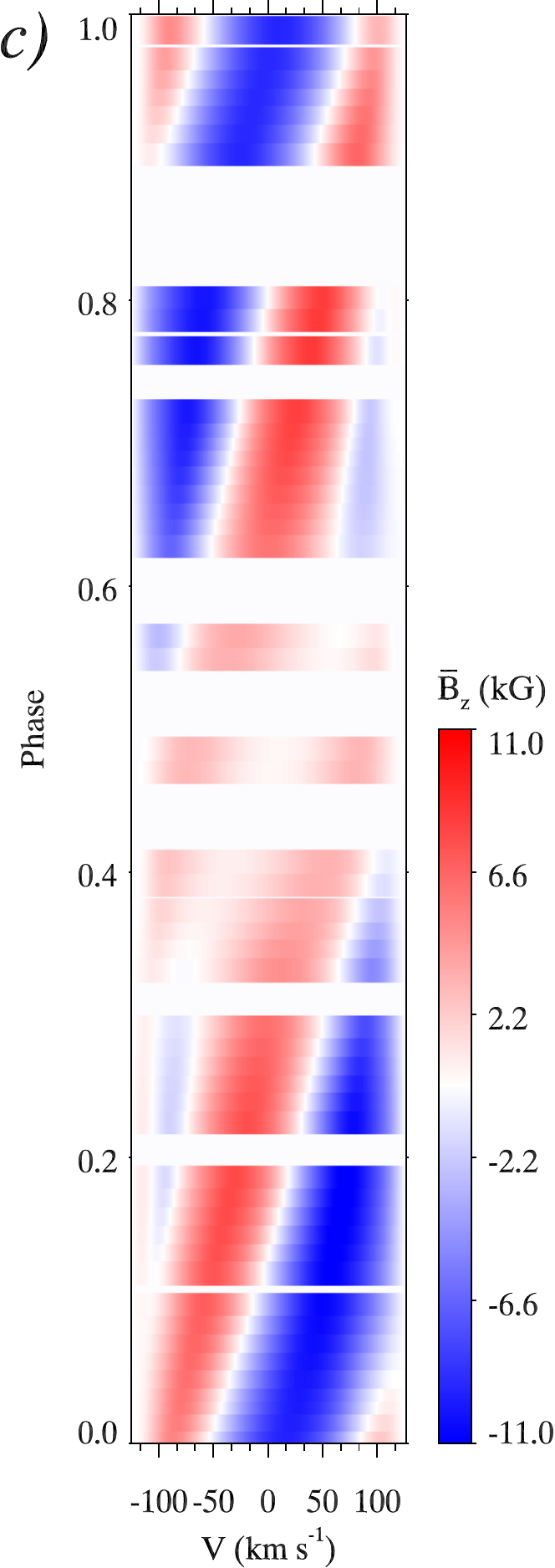}
\caption{Cumulative Stokes $V$ spectra corresponding to {\bf a)} the observed Fe LSD profiles, {\bf b)} the observed Si LSD profiles and {\bf c)} the magnetic field geometry of \hr\ reported by \citet{bailey:2012}.
}
\label{fig:prf_csv}
\end{figure}

The cumulative Stokes $V$ (CSV) profile diagnostic technique was suggested by \citet{gayley:2014} and further developed by \citet{kochukhov:2015a}. In this method the observed circular polarisation profiles of individual spectral lines or multi-line LSD profiles are converted to a spatially resolved measure of the longitudinal magnetic field density $\overline{B}_{\rm z}$, which essentially represents a Doppler-resolved equivalent of the widely used mean line-of-sight (longitudinal) magnetic field \bz. In comparison to \bz, which is a scalar quantity characterising the longitudinal magnetic field averaged over the entire stellar disk, $\overline{B}_{\rm z}$ is a velocity-dependent measure of the line-of-sight field component averaged over the stripes of constant Doppler shift. Consequently, $\overline{B}_{\rm z}$ is less affected by cancellation of polarisation signals coming from regions with opposite field polarities and can provide a meaningful measure of the longitudinal field at crossover phases, when \bz\ is zero. An analysis of CSV profiles allows one to qualitatively assess the topology of stellar magnetic field and to make an estimate of the local magnetic field strength without the need of sophisticated line profile modelling such as MDI. The CSV method is particularly suitable for fast-rotating stars suspected of hosting complex magnetic field geometries.

Here we have applied the CSV diagnostic procedure to the Si and Fe LSD Stokes $V$ profiles of \hr\ and compared the resulting $\overline{B}_{\rm z}$ profiles to the predictions of the magnetic field geometry model derived by \citet{bailey:2012}. According to this study, the field of \hr\ is best approximated as a superposition of the aligned dipolar $B_{\rm d}=-9.6$~kG, quadrupolar $B_{\rm d}=-23.2$~kG and octupolar $B_{\rm oct}=1.9$~kG components, along with $i=55\degr$ and $\beta=78\degr$.

The observed CSV profiles were obtained by a weighted average of the blue-to-red and red-to-blue integrals over the Stokes $V$ profiles, normalised by the corresponding residual Stokes $I$ profiles (see \citealt{kochukhov:2015a} for details). The resulting dynamic CSV spectra are presented in the panels {\bf a)} and {\bf b)} of Fig.~\ref{fig:prf_csv}. Each panel in this figure shows a phase variation (vertical axis) of the longitudinal field density (colour scale) spatially resolved in the direction (horizontal axis) perpendicular to the stellar rotational axis.

One can infer from Fig.~\ref{fig:prf_csv} that the Si and Fe CSV profiles of \hr\ are very similar despite a noticeable difference in the morphology of the underlying LSD Stokes $V$ profiles (modelled in Sect.~\ref{mdi}). The CSV spectra of both chemical elements indicate a field topology with an extended region of weaker positive magnetic field (longitudinal field density $\overline{B}_z \le 2.5$~kG) and a smaller region of stronger negative field ($\overline{B}_z \ge -5.4$~kG). The CSV profiles are, therefore, generally compatible with a somewhat distorted and asymmetric dipolar field structure but do not reveal any evidence of a quadrupole-dominated field geometry.

The observed profiles can be compared with the theoretical CSV spectra obtained by integrating the line-of-sight component of the field geometry model by \citet{bailey:2012} for the same set of rotational phases as sampled by our spectropolarimetric observations. These theoretical CSV profiles, shown in Fig.~\ref{fig:prf_csv}c, are morphologically more complex than the observed ones. For example, the secondary negative magnetic field region predicted by the model is not observed. Moreover, this field topology predicts a much stronger overall longitudinal magnetic field density ($\overline{B}_z$ ranging from $-11.3$ to 8.5~kG) than is evident from the observed CSV data. Thus, prior to any MDI inversions, we are able to conclude based on the CSV analysis results that, although the field geometry proposed by \citet{bailey:2012} successfully describes the observed phase variation of the disk-integrated longitudinal field, it is, in fact incompatible with the observed Stokes $V$ profiles and appears to significantly overestimate both the strength and the degree of complexity of the surface magnetic field in \hr. These qualitative conclusions fully agree with an outcome of the MDI modelling presented below.

\subsection{Magnetic Doppler imaging}
\label{mdi}

\subsubsection{Magnetic topology and distributions of Si, Cr and Fe}

\begin{figure}[!th]
\centering
\figps{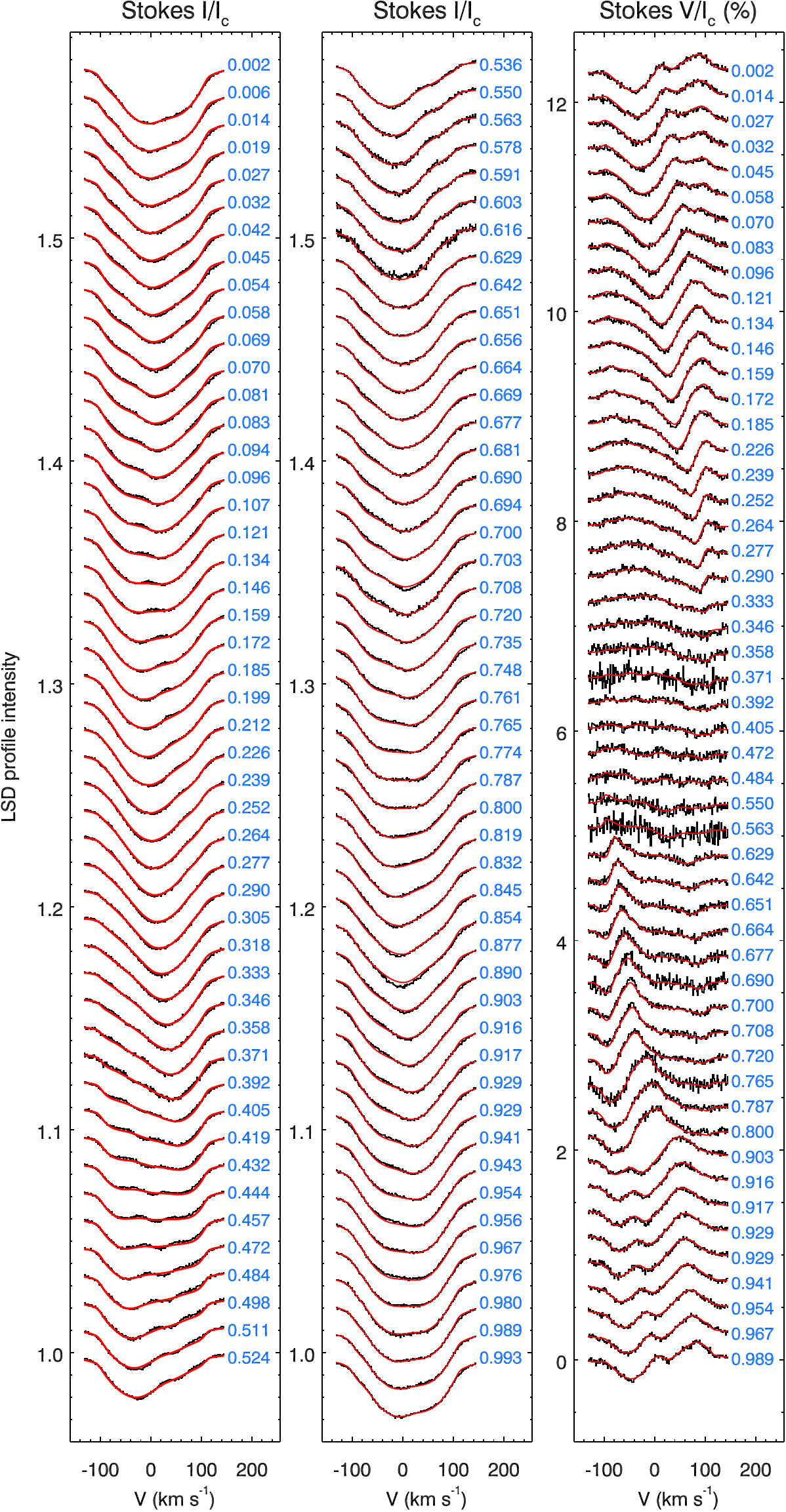}
\caption{Comparison of the observed Si LSD Stokes $I$ and $V$ profiles of \hr\ with the fit achieved by the magnetic inversion code. Observations are shown with black histograms. Calculations for the final magnetic and chemical spot maps are shown with the solid red lines. Spectra corresponding to different rotation phases are offset vertically. Rotation phases are indicated to the right of each spectrum.}
\label{fig:prf_lsd_si}
\end{figure}

\begin{figure}[!th]
\centering
\figps{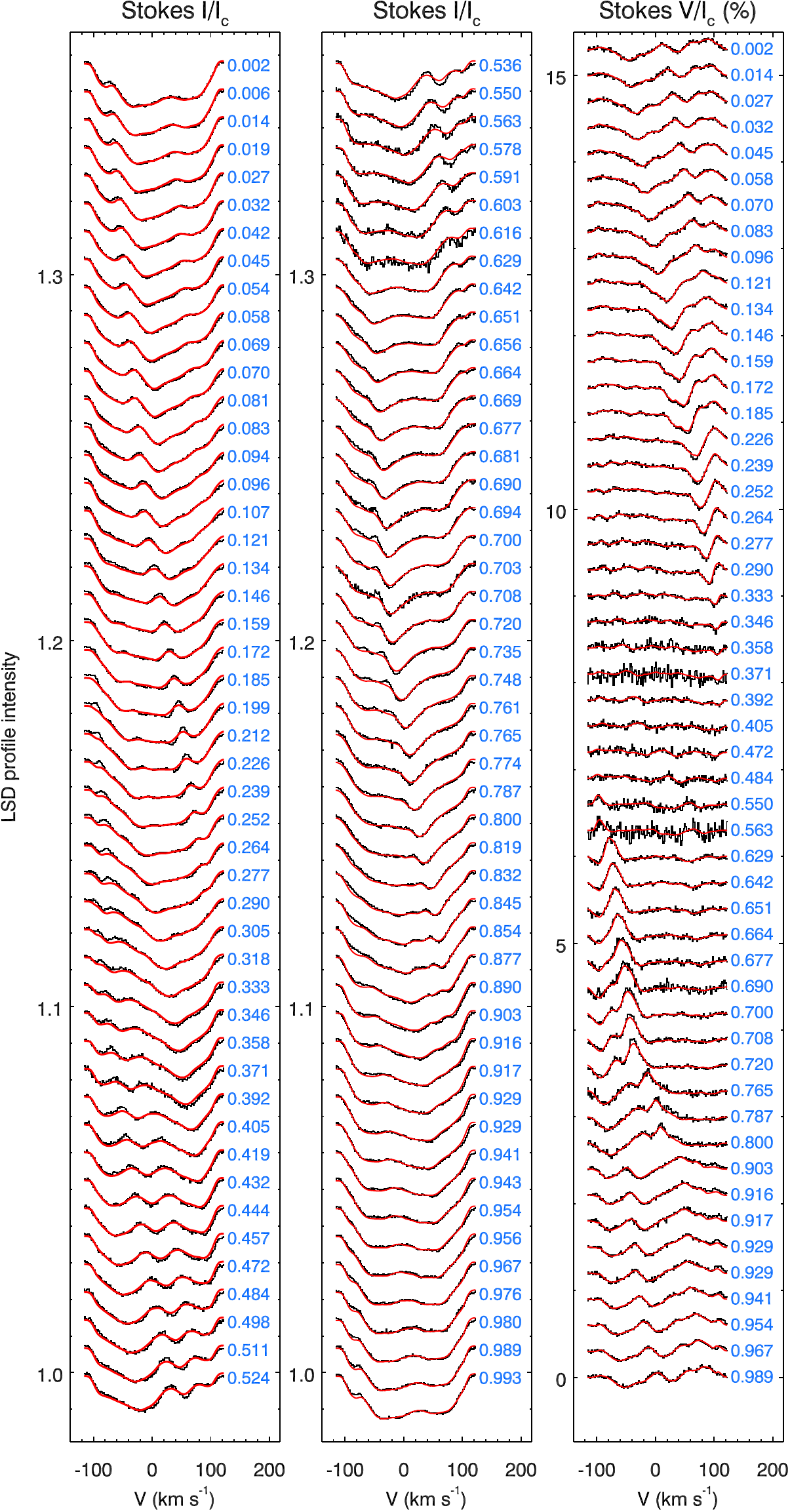}
\caption{Same as Fig.~\ref{fig:prf_lsd_si} for Cr LSD Stokes $I$ and $V$ profiles.}
\label{fig:prf_lsd_cr}
\end{figure}

\begin{figure}[!th]
\centering
\figps{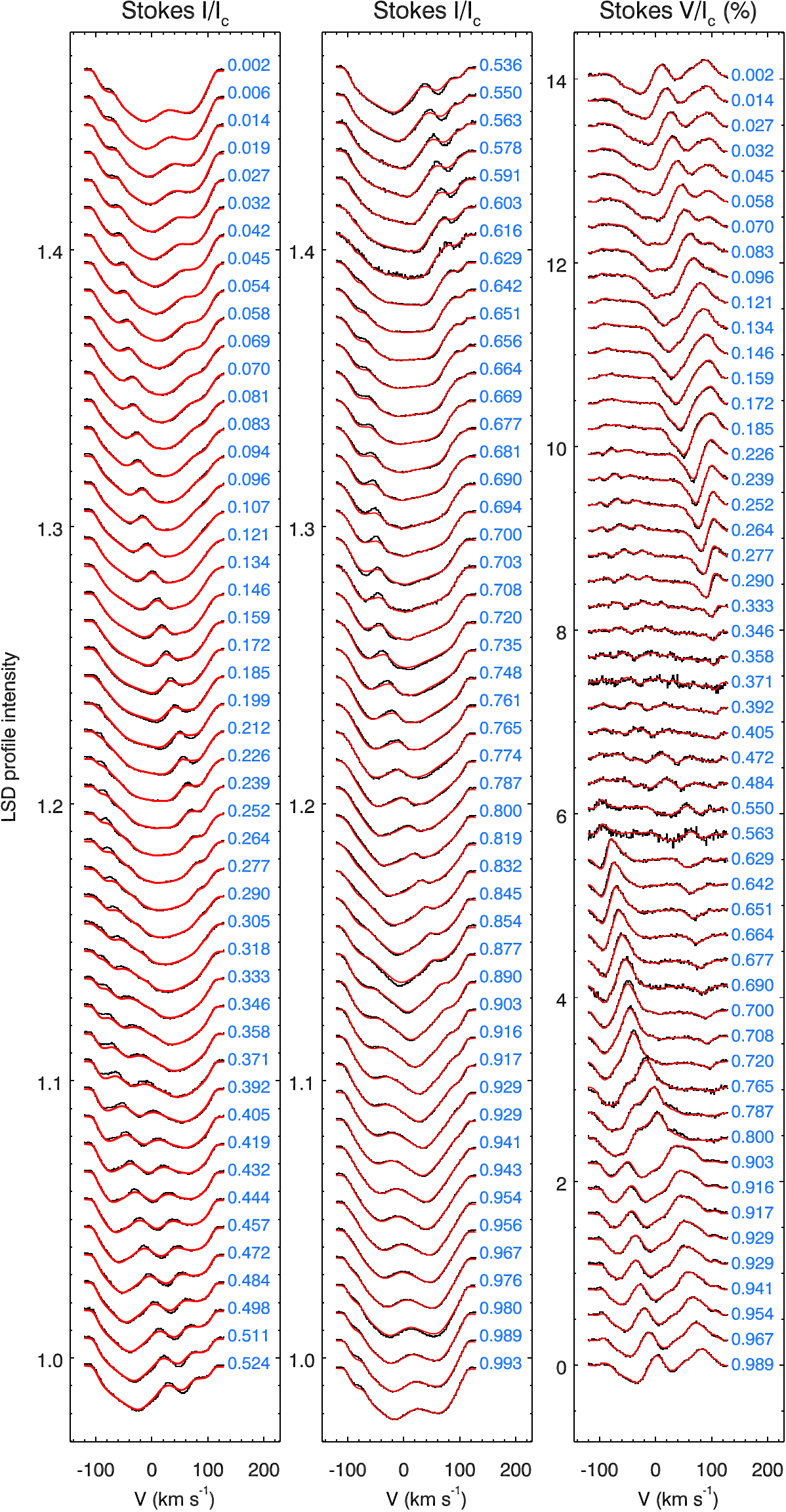}
\caption{Same as Fig.~\ref{fig:prf_lsd_si} for Fe LSD Stokes $I$ and $V$ profiles.}
\label{fig:prf_lsd_fe}
\end{figure}

\begin{figure*}[!th]
\centering
\firrps{15.3cm}{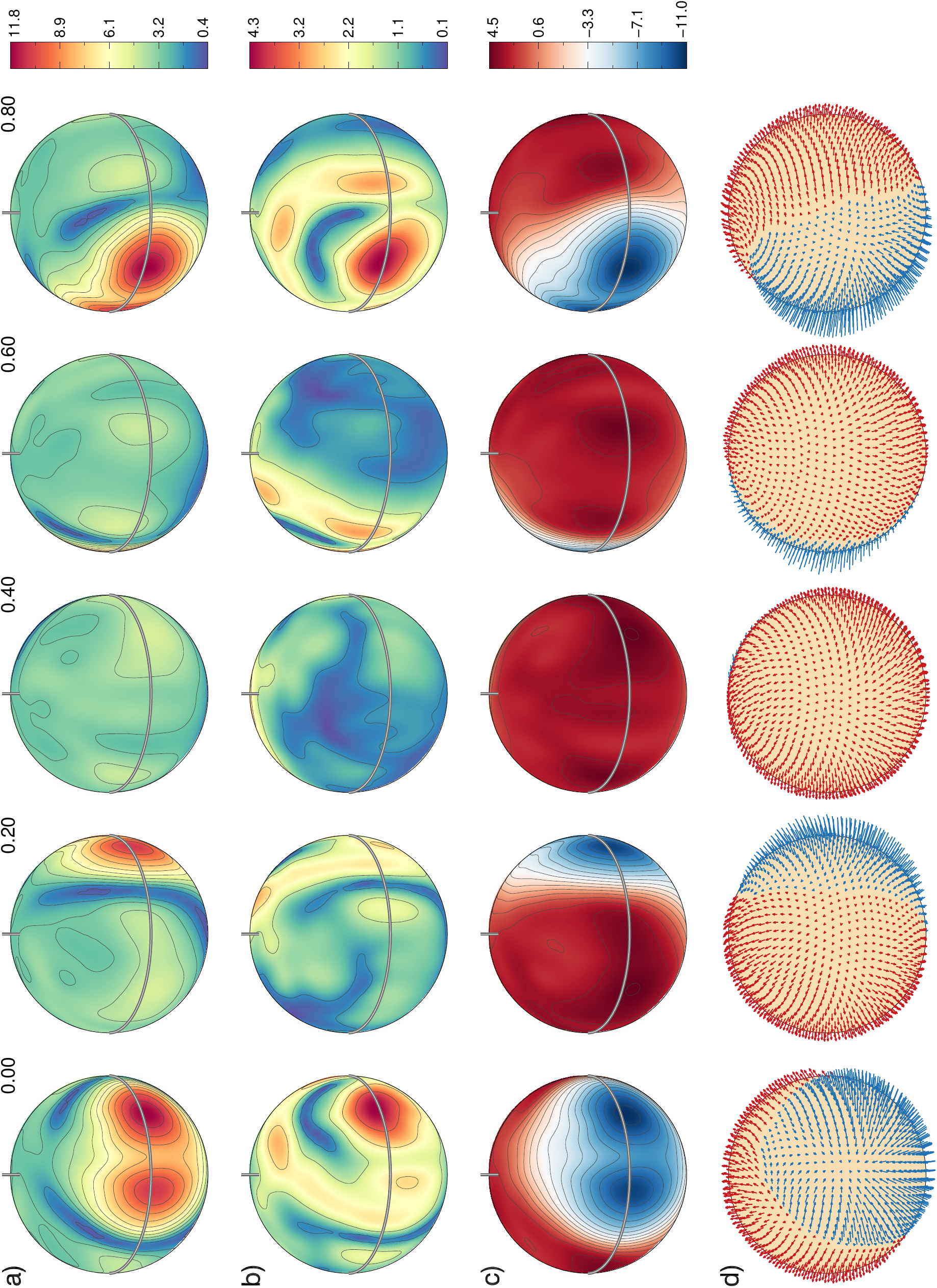}
\caption{Surface magnetic field distribution of \hr\ derived from the Si LSD profiles. The star is shown at five rotation phases, which are indicated above each spherical plot column. The inclination angle is $i=65\degr$. The spherical plots show the maps of {\bf a)} field modulus, {\bf b)} horizontal field, {\bf c)} radial field, and {\bf d)} field orientation. The contours over spherical maps are plotted with a step of 1~kG. The thick line and the vertical bar indicate positions of the rotational equator and the pole, respectively. The colour bars give the field strength in kG. The two different colours in the field orientation map correspond to the field vectors directed outwards (red) and inwards (blue).}
\label{fig:fld_si}
\end{figure*}

\begin{figure*}[!th]
\centering
\firrps{15.3cm}{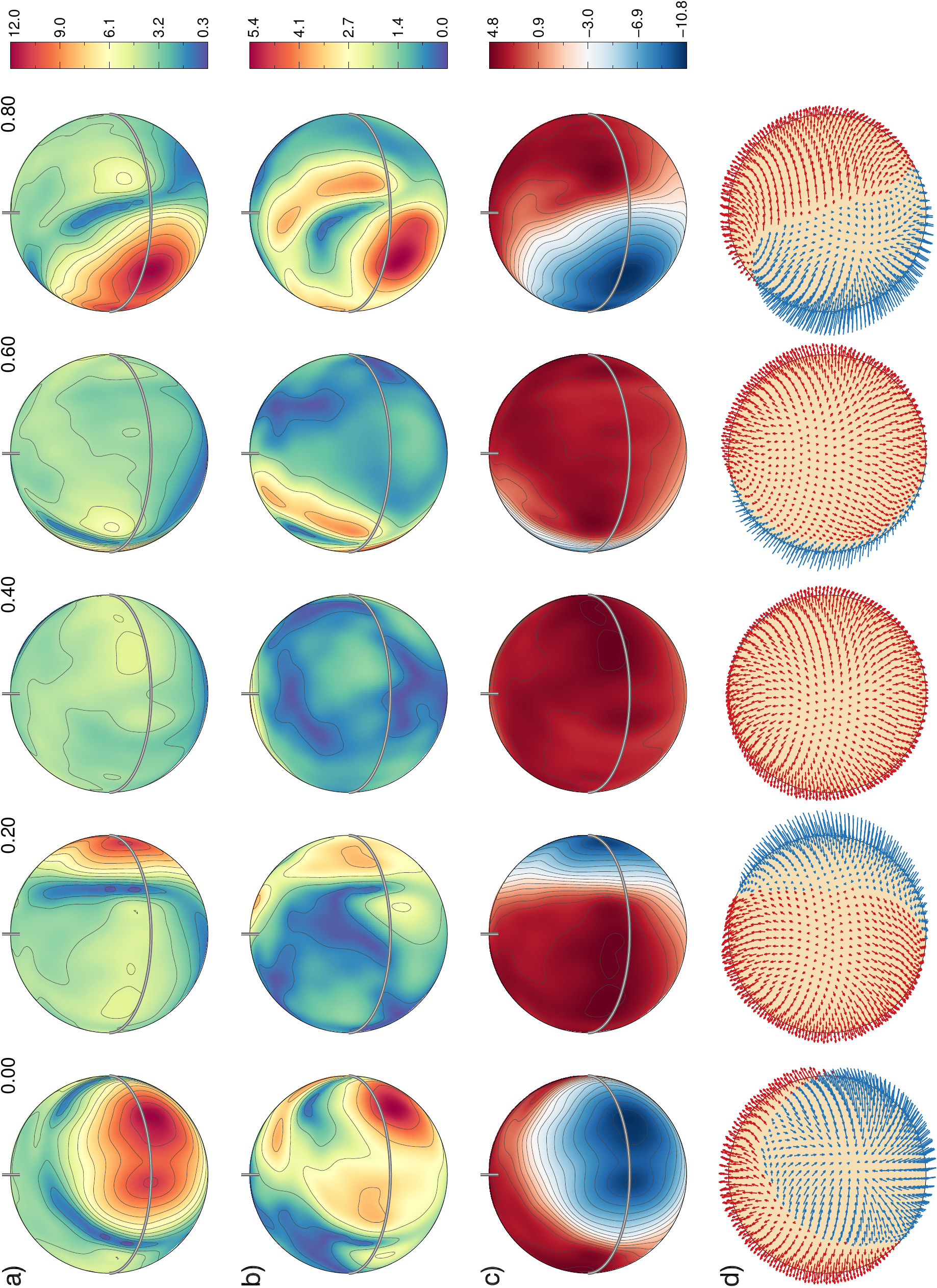}
\caption{Same as Fig.~\ref{fig:fld_si} for the surface magnetic field distribution derived from the Cr LSD profiles.}
\label{fig:fld_cr}
\end{figure*}

\begin{figure*}[!th]
\centering
\firrps{15.3cm}{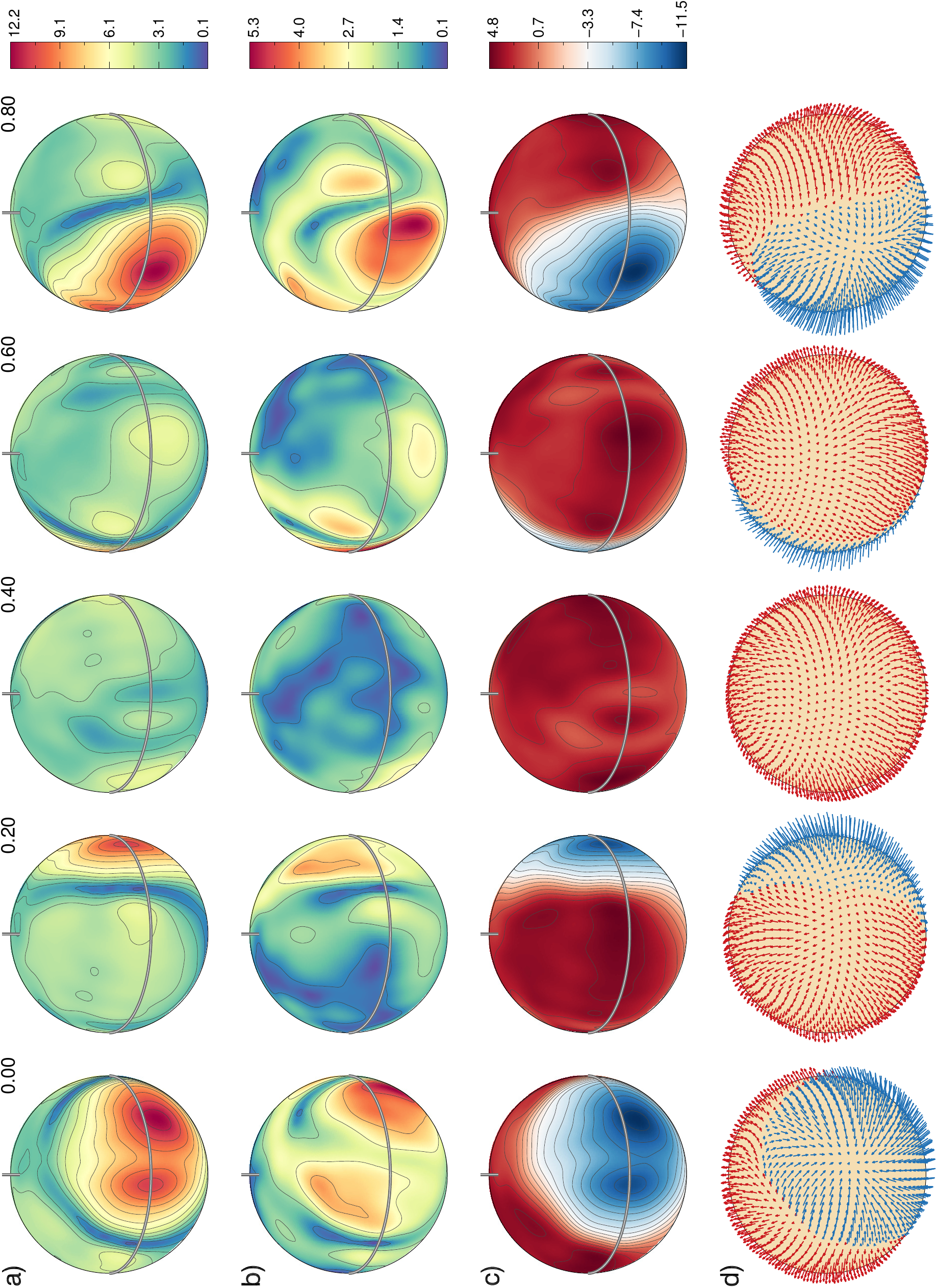}
\caption{Same as Fig.~\ref{fig:fld_si} for the surface magnetic field distribution derived from the Fe LSD profiles.}
\label{fig:fld_fe}
\end{figure*}

\begin{figure*}[!th]
\centering
\firrps{15.3cm}{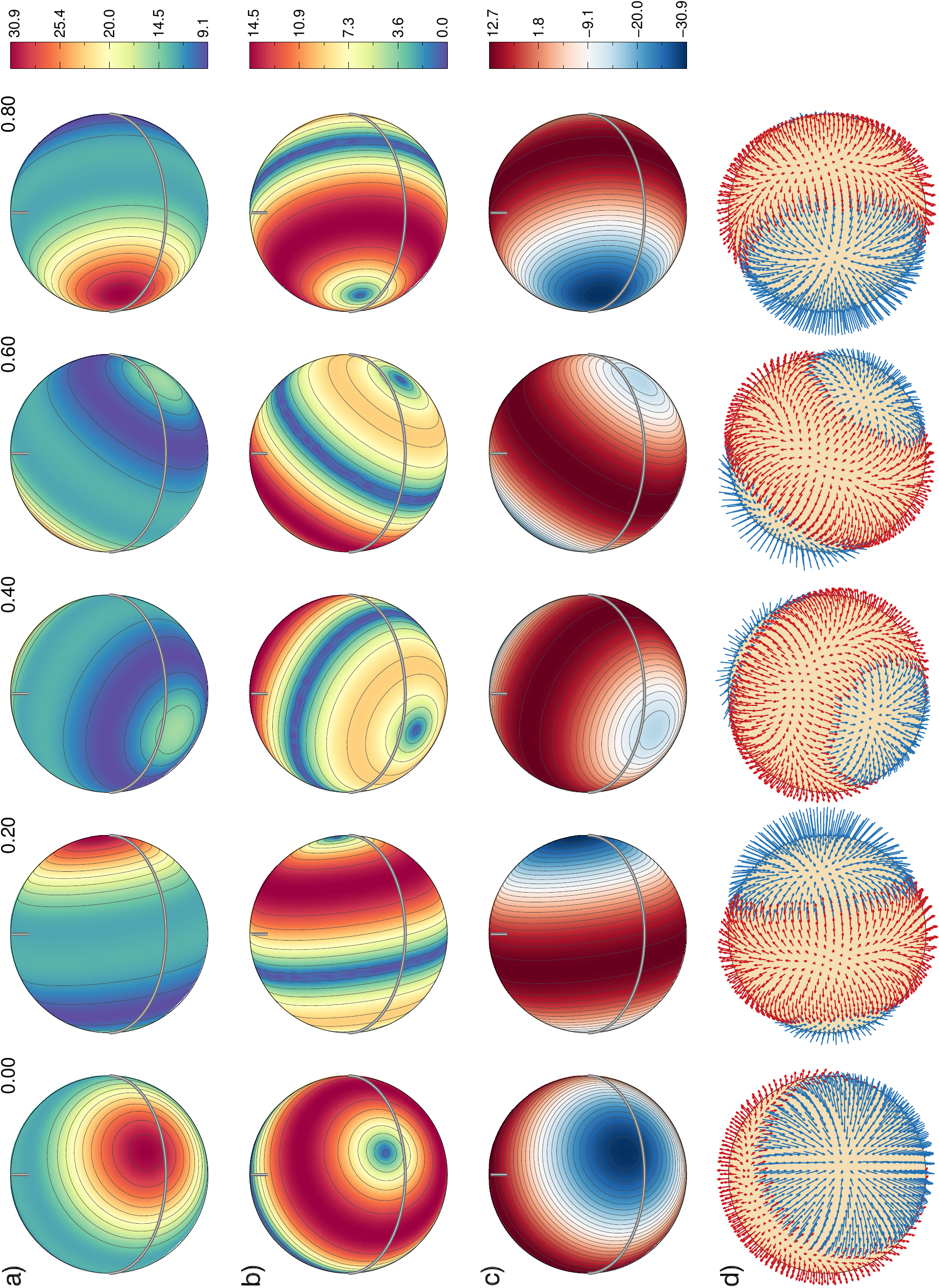}
\caption{Same as Fig.~\ref{fig:fld_fe} for the surface magnetic field distribution derived by \citet{bailey:2012}. The contours over spherical maps are plotted with a 2~kG step.}
\label{fig:fld_jb}
\end{figure*}

\begin{figure}[!th]
\centering
\figps{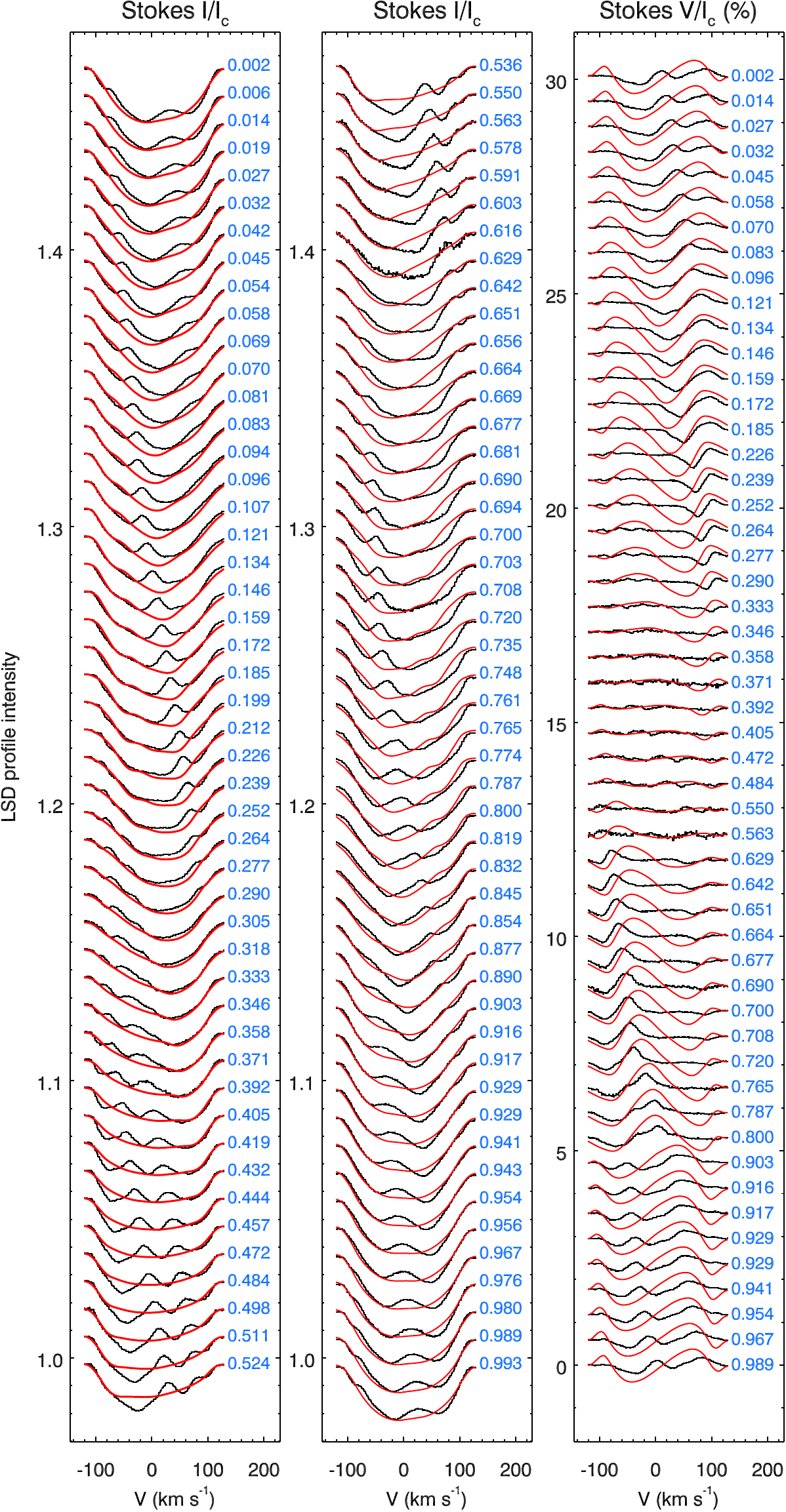}
\caption{Same as Fig.~\ref{fig:prf_lsd_si} for the comparison of the observed Fe LSD Stokes $I$ and $V$ profiles and forward calculations with the iron abundance distribution and magnetic field geometry of \citet{bailey:2012}.}
\label{fig:prf_lsd_fe_jb}
\end{figure}

We have reconstructed the magnetic field topology of \hr\ by modelling the Si, Cr, and Fe LSD profiles with the help of the {\sc InversLSD} MDI code described by \citet{kochukhov:2014}. This inversion code performs simultaneous mapping of a vector magnetic field and one additional scalar parameter, in this case chemical abundance, based on an interpolation within pre-tabulated grids of theoretical local Stokes LSD profiles. Following the approach previously used for the magnetic mapping of CU\,Vir \citep{kochukhov:2014}, the local Stokes LSD profiles of \hr\ were computed by applying the least-squares deconvolution procedure to a set of local synthetic four Stokes parameter spectra covering the entire observed wavelength range and including all relevant absorption lines (those comprising the Si, Cr, and Fe LSD masks as well as all significant blends). These theoretical spectra were tabulated for a grid of 15 limb angles, 31 field strength values between 0 and 30~kG, 15 field vector orientations with respect to the line of sight, and 13--15 element abundance values. The element abundance variations were treated with the help of three separate grids of {\sc LLmodels} atmospheres, calculated with the parameters determined above and changing logarithmic abundances of either Si, Cr, or Fe by 0.25~dex within the full range necessary for the respective inversions. This allowed us to take into account modifications of the local atmospheric structure and non-uniform continuum brightness (although according to \citet{lehmann:2007}, \citet{kochukhov:2012} and \citet{kochukhov:2017} these effects usually have only a marginal influence on the stellar surface maps).

The magnetic field geometry of \hr\ was parameterised with a spherical harmonic expansion including modes with an angular degree up to $\ell_{\rm max}=20$. This value was chosen to approximately match the theoretical spatial resolution enabled by the \vs\ of \hr\ \citep{fares:2012}. Both the poloidal and toroidal terms were included in the expansion. The inversions were initiated with zero magnetic field and a homogeneous abundance distribution. The regularisation parameters, controlling the strength of the harmonic regularisation for the magnetic field map and the Tikhonov regularisation for abundance maps, were adjusted as described by \citet{kochukhov:2017}. A series of MDI inversions was carried out for different values of the projected rotational velocity \vs\ and inclination angle $i$, aiming to find optimal values of these parameters. This analysis yielded \vs\,=\,$106\pm2$~\kms\ and $i=65\pm10\degr$, both of which agree reasonably well with the determinations by \citet{bailey:2012}.

\begin{figure*}[!t]
\centering
\fifps{5.9cm}{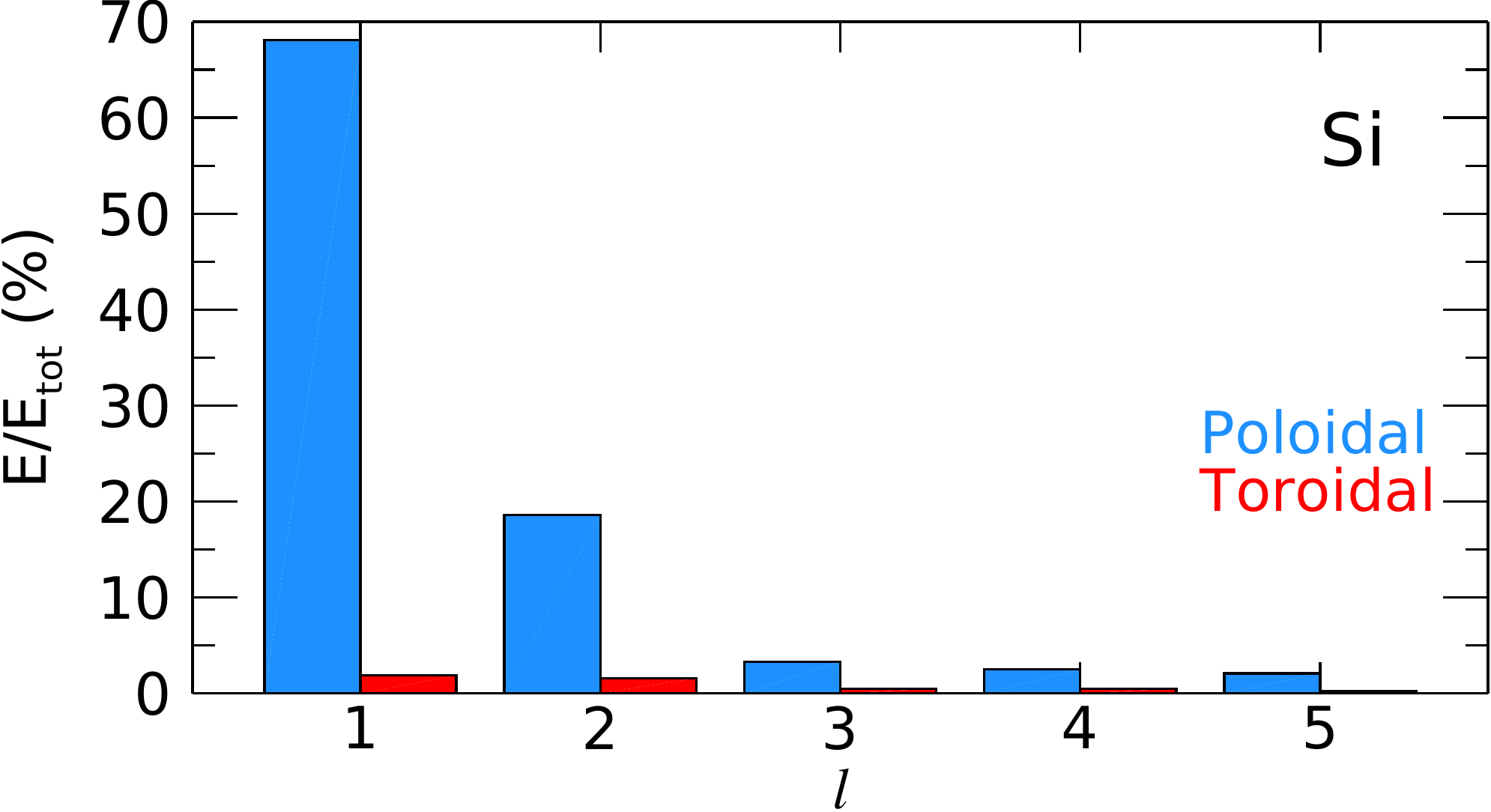}\hspace*{0.1cm}
\fifps{5.9cm}{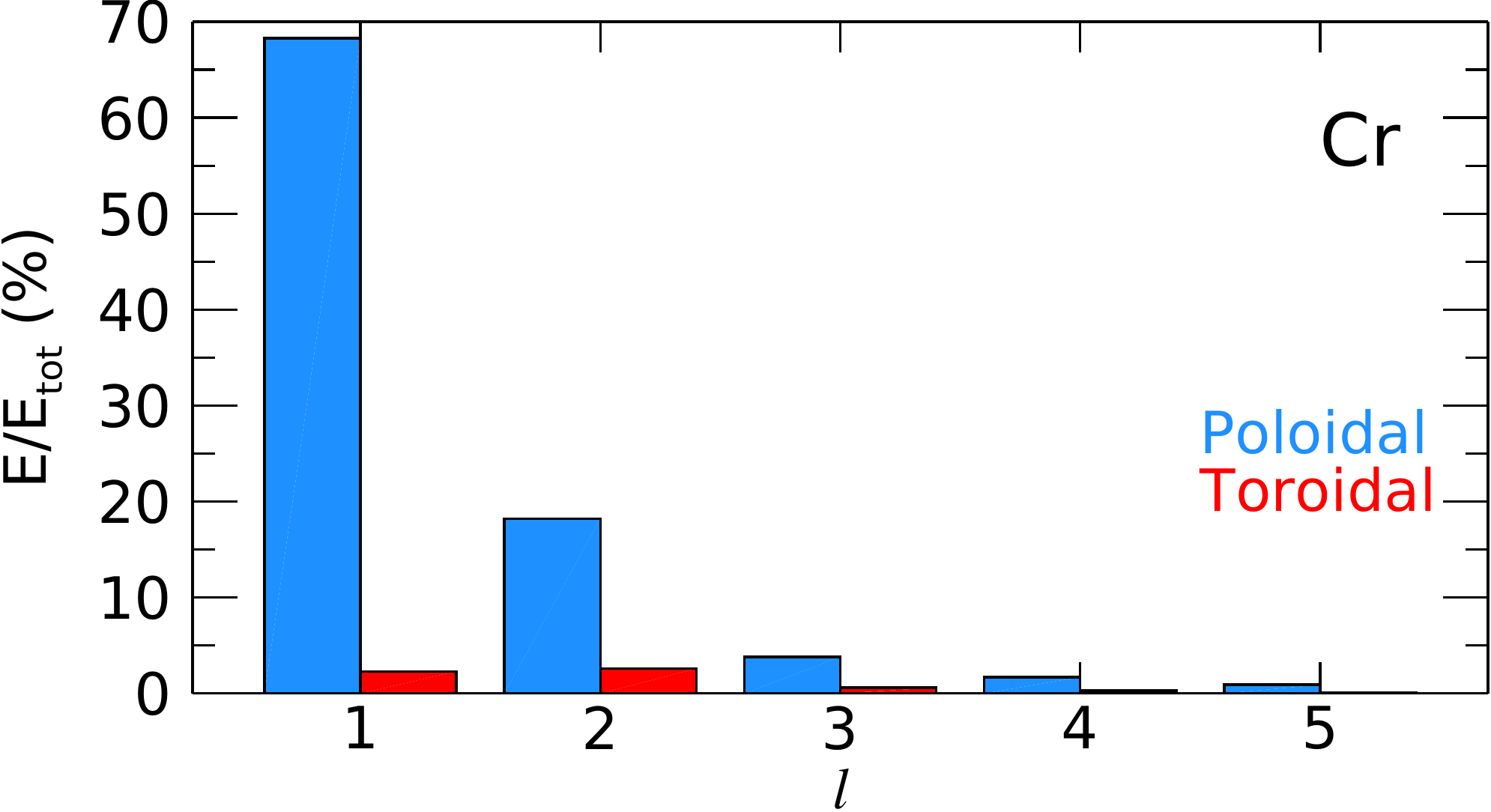}\hspace*{0.1cm}
\fifps{5.9cm}{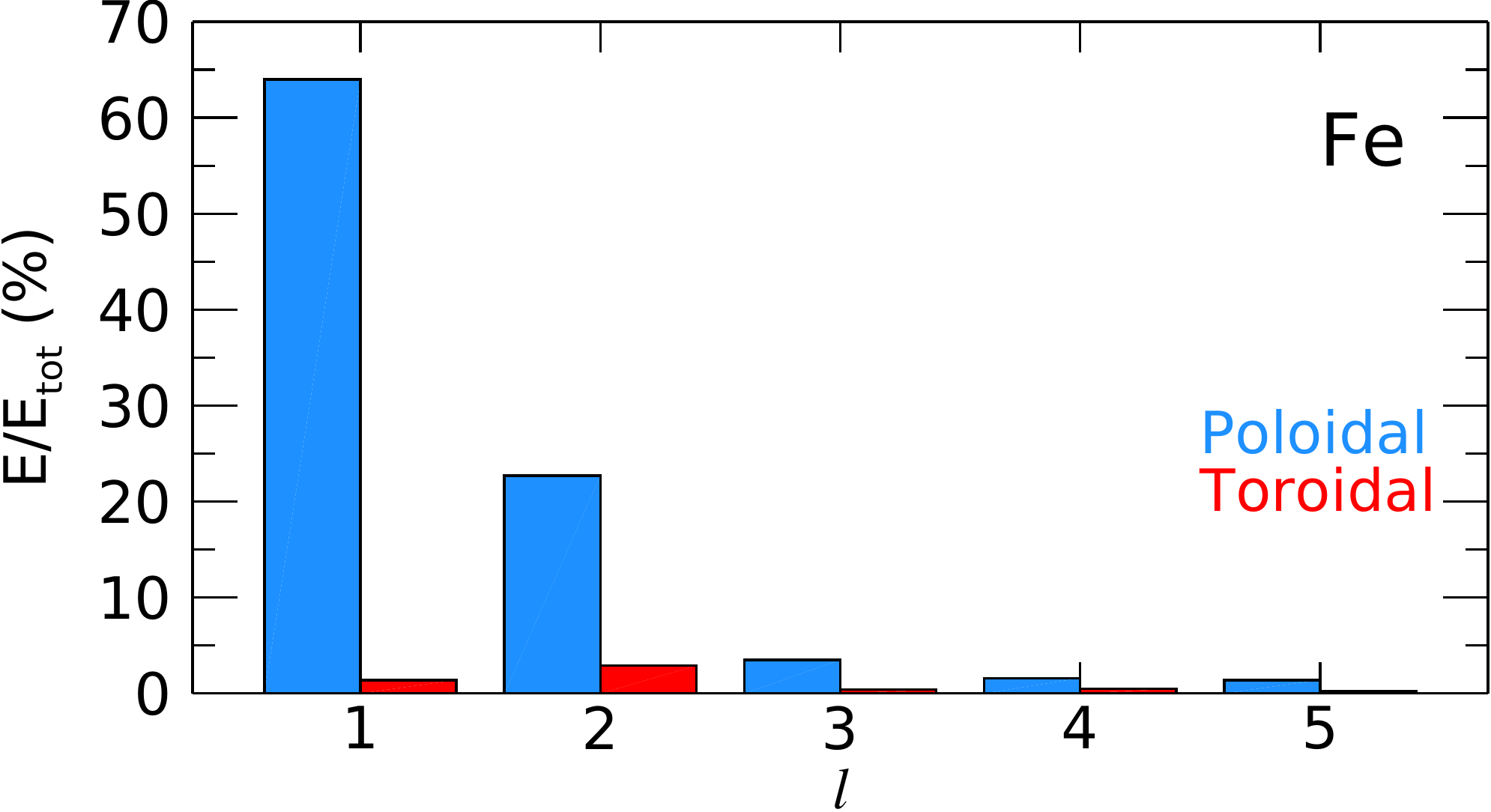}
\caption{Relative energies of the poloidal and toroidal harmonic modes with different angular degrees $\ell$ for the magnetic field topology of \hr\ derived from Si (left), Cr (middle) and Fe (right) LSD profiles.}
\label{fig:fld_coef}
\end{figure*}

The final fits by {\sc InversLSD} to the observed Stokes $I$ and $V$ LSD profiles of Si, Cr, and Fe are presented in Figs.~\ref{fig:prf_lsd_si}--\ref{fig:prf_lsd_fe}. The corresponding spherical plots of the total magnetic field strength, horizontal field, radial field and the field vector orientation are shown in Figs.~\ref{fig:fld_si}--\ref{fig:fld_fe}. For comparison, we also illustrate in Fig.~\ref{fig:fld_jb} the magnetic geometry model obtained in the previous detailed study of \hr\ by \citet{bailey:2012}. Finally, the harmonic energy distribution of the Si, Cr, and Fe-based magnetic field maps is schematically illustrated in Fig.~\ref{fig:fld_coef} for modes up to $\ell=5$ (the contribution of higher order terms is negligible). Different statistical characteristics of these field geometries are summarised in Table~\ref{tbl:fld}.

Turning attention to the magnetic field maps shown in Figs.~\ref{fig:fld_si}--\ref{fig:fld_fe}, it is evident that the best-fitting magnetic field distribution is basically dipolar. However, this dipole is strongly asymmetric and distorted, with a much stronger and smaller negative magnetic pole where the radial field reaches $-11$~kG and a weaker, more extended positive magnetic field region with a radial field of up to 4.5--4.8~kG. The global mean of the field strength is found to be 4.0--4.4~kG. The maximum local field modulus is about 12~kG in all three magnetic field maps. The corresponding mean field modulus is predicted to vary between 3.4 and 6.1~kG, with a sharp maximum at phase 0.95 and a broad minimum in the phase interval 0.25--0.65. The phase-averaged value of the field modulus is predicted to be 4.1--4.6~kG.

The details of the magnetic field topologies inferred from the independent LSD profile modelling of Si, Cr, and Fe agree very well, with typical local standard deviation of only 0.3--0.4~kG for all three magnetic field vector components and the field modulus. This concordance is particularly noteworthy given the somewhat different appearance of the Fe and Cr Stokes $V$ LSD profiles (Figs.~\ref{fig:prf_lsd_cr} and \ref{fig:prf_lsd_fe}) and the significantly simpler morphology of the Si Stokes $V$ LSD profiles seen in Fig.~\ref{fig:prf_lsd_si}. The latter appears to be a consequence of a less structured Si abundance distribution. All three magnetic field maps indicate that the negative pole is formed by two magnetic spots of unequal strength, with the trailing spot located at a longitude of about 45\degr\ and including a substantial horizontal magnetic component.

Comparing our MDI magnetic field maps with those of the quadrupole-dominated parametric magnetic field geometry ($B_{\rm d}=-9.6$~kG, $B_{\rm d}=-23.2$~kG $B_{\rm oct}=1.9$~kG, $i=55\degr$, $\beta=78\degr$) proposed by \citet{bailey:2012}, we find that our inversions favour much weaker global field strengths. In Fig.~\ref{fig:prf_lsd_fe_jb} we show forward calculations of $I$ and $V$ profiles based on the \citeauthor{bailey:2012} model. Although the $V$ profiles derived from the quadrupole-dominated model show some qualitative resemblance to the observed profiles, they are systematically less structured and have amplitudes which are typically too large by factors of order 2. On the other hand, although the $I$ line profiles resulting from the quadrupole-dominated model do not reproduce the detailed shape of the observations, they generally have about the right depth and show a reasonable resemblance to the observations. This better accord is probably due to the fact that the I profiles are determined mainly by the adopted abundance distributions  (particularly of Fe) rather than the magnetic field. 

The analysis of the harmonic energy distribution of the Si, Cr, and Fe magnetic maps (Fig.~\ref{fig:fld_coef}) reveals the dominant contribution (65--71\% in terms of the total magnetic field energy) of the dipolar component inclined by about 90\degr\ with respect to the stellar rotational axis. All quadrupolar ($\ell=2$) modes together contribute 20--26\% of the magnetic energy. Contributions of higher order modes with $\ell\ge 3$ do not exceed 4.5\%. No evidence of a substantial toroidal field is found in our magnetic inversion results.

\begin{figure*}[!th]
\centering
\firrps{15cm}{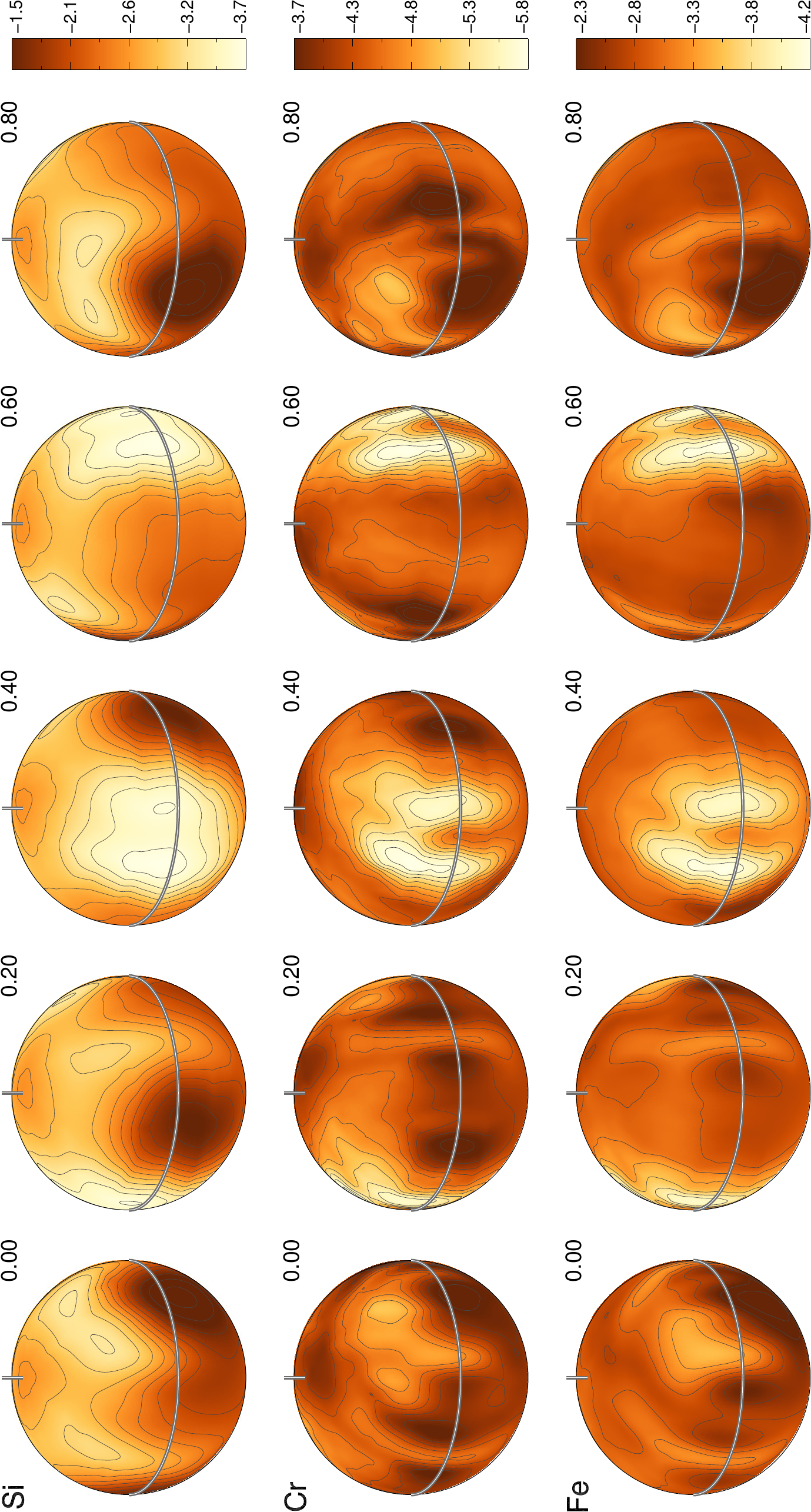}
\caption{Silicon, chromium, and iron surface abundance distributions reconstructed from the LSD profiles of these elements simultaneously with the magnetic field mapping. The star is shown at five rotational phases, as indicated next to each plot. The contours over spherical maps are plotted with a 0.2~dex step. The side bars give element abundances in logarithmic units $\log N_{\rm el}/N_{\rm tot}$.}
\label{fig:abn_lsd}
\end{figure*}

The Si, Cr, and Fe abundance distributions, reconstructed simultaneously with the magnetic field maps, are presented in Fig.~\ref{fig:abn_lsd}. The chemical maps of these elements differ in detail but also have several properties in common. For instance, the contrast is about 2~dex in all three cases. Also, all three maps exhibit overabundance features coinciding with the stronger of the two magnetic spots at the negative pole as well as zones of relative element underabundance in the vicinity of the extended positive pole. For iron and chromium, this underabundance structure splits into two distinct narrow bands extended in latitude, possibly forming a ring centred at around longitude 190\degr. The effect of this underabundance ring is readily evident in the highly structured Cr and Fe Stokes $I$ LSD profiles between phases 0.4 and 0.6. The Cr and Fe distributions are generally similar to each other. However, the chromium map shows several smaller overabundance spots at the rotational equator not present in the Fe map.

\subsubsection{Distributions of Mg, Ti and Nd}

\begin{table*}[!th]
\centering
\caption{Statistical characteristics of the magnetic field maps reconstructed from Si, Cr, and Fe LSD profiles of \hr. 
\label{tbl:fld}}
\begin{tabular}{l|cc|cc|ccc}
\hline\hline
& \multicolumn{2}{c|}{Local field strength (kG)} & \multicolumn{2}{c|}{Mean field modulus (kG)} & $E_{\ell=1}/E_{\rm tot}$ & $E_{\ell=2}/E_{\rm tot}$ & $E_{\rm pol}/E_{\rm tot}$ \\
Element & Mean & Range & Mean & Range & (\%) & (\%) & (\%) \\
\hline
Si      &   3.95  &  0.44--11.72   &      4.06 & 3.19--5.69  &       70.0        &     20.1  &       95.3 \\
Cr       &  4.36  &  0.28--11.95   &      4.57 & 3.56--6.37   &      70.6     &        20.9    &     93.8 \\
Fe       & 4.27  &  0.45--12.14    &     4.38 & 3.34--6.20     &    65.4   &          25.6      &   94.4 \\
\hline
\end{tabular}
\end{table*}

\begin{figure*}[!th]
\centering
\firrps{15cm}{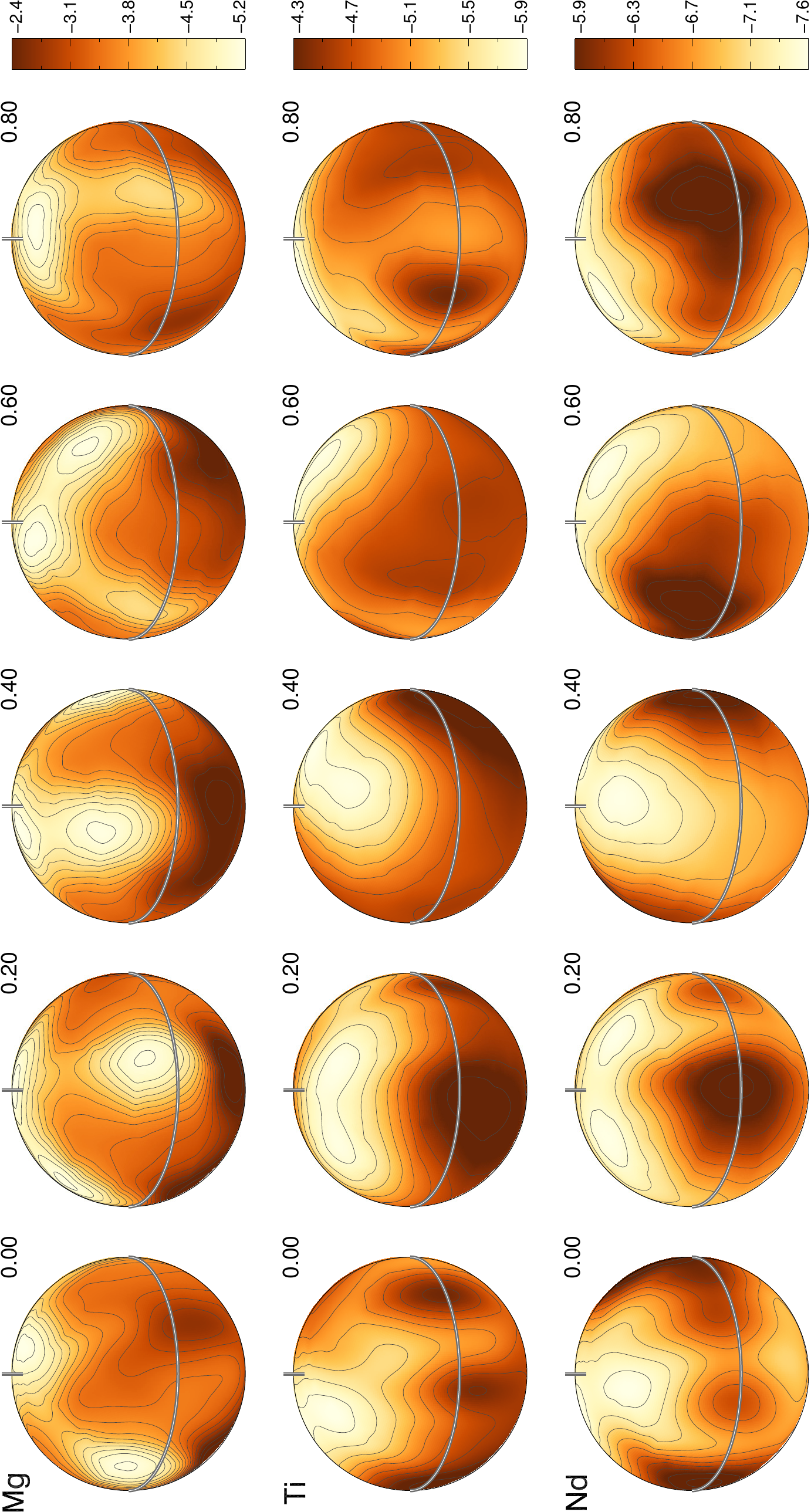}
\caption{Same as Fig.~\ref{fig:abn_lsd} for magnesium, titanium, and neodymium surface abundance distributions reconstructed from individual spectral lines of these elements.}
\label{fig:abn}
\end{figure*}

In addition to the chemical abundance maps of Si, Cr, and Fe recovered from the Stokes $I$ and $V$ LSD profiles as part of the MDI inversions, we also studied the surface distributions of several other elements using intensity profiles of individual spectral lines. Due to the rapid rotation of \hr, it is difficult to find unblended lines of elements other than Si, Cr, Fe. Nevertheless, surface distributions of Mg, Ti, and Nd could be derived by modelling the 448.1, 480.5, and 494.3~nm lines of these elements respectively. Chemical abundance maps were obtained using {\sc Invers10} \citep{piskunov:2002a,kochukhov:2002c} and adopting the average of the magnetic field topologies determined above. In each case the line list included up to 10 blending lines in addition to the main spectral features. Most of the secondary lines belonged to Fe and Cr, allowing the treatment of inhomogeneous distributions of these elements according to the previously derived abundance maps.

The resulting surface distributions of Mg, Ti, and Nd are presented in Fig.~\ref{fig:abn}. All three elements exhibit horizontal inhomogeneities with the abundance contrast of 1.6--2.8~dex. Some of these surface distributions show certain similarities to the Si, Cr, and Fe abundance maps recovered from the LSD profiles. The distribution of Nd is dominated by two distinct overabundance areas, seen at the rotational phases 0.2 and 0.8, which are offset from the magnetic equator by $\sim$\,30\degr\ in longitude towards the positive magnetic field regions. Curiously, the Si map is also dominated by two spots, with one of them almost coinciding with the spot of Nd (phase 0.2), but another one offset by nearly 60\degr\ from the second Nd spot (phase 0.8). The region with the minimum element abundance in the vicinity of the extended, weak-field, positive magnetic pole occurs, in one form or another, in every abundance map. On the other hand, with the exception of Nd, no such region is seen close to the strong-field, negative magnetic pole.

\section{Discussion}
\label{disc}

\subsection{Magnetic field topology of \hr}

In this study we have carried out the first detailed polarisation line profile modelling of the young Bp star \hr, a star previously suggested to host an unusual, predominantly quadrupolar, very strong surface magnetic field \citep{landstreet:1990,bailey:2012}. Contrary to these predictions, our MDI inversions reveal a much weaker magnetic field, which to a first approximation resembles a distorted, offset dipole. This field topology still resembles a nearly aligned superposition of a dipole and a linear quadrupole. Moreover, our finding that the negative magnetic pole is comprised of a pair of high-contrast magnetic features qualitatively agrees with the conjecture of previous quadrupolar models, which predicted a strong magnetic spot at the same location. These results appear to reinforce the notion that an aligned dipole plus quadrupole magnetic field parameterisation provides a satisfactory first-order description of Ap-star magnetic fields and is useful for coarse analyses of large stellar samples \citep[e.g.][]{landstreet:2000}. 

However, despite this qualitative agreement with our detailed MDI maps, other predictions of the parametric field topology models are entirely spurious. In the specific case of \hr, quantitative MDI results differ markedly from the corresponding quadrupolar model predictions. MDI maps indicate that the dipolar component dominates the stellar field topology, contributing 65--70\% of the total magnetic field energy, while all quadrupolar ($\ell=2$) modes together provide a factor of 2.6--3.5 less contribution to the magnetic energy budget. This is a much larger dipolar contribution compared to the findings of previous field topology studies by \citet{landstreet:1990} and \citet{bailey:2012}, both of which adopted a combination of aligned dipole, linear quadrupole and linear octupole. The first paper found a fairly wide variety of parameter combinations that could explain the observed \bz\ variations, with models based on a realistic limb-darkening function and (arbitrarily assumed) $i=90\degr$ requiring a quadrupole-to-dipole polar field strength ratio of $B_{\rm q}/B_{\rm d}$\,$\sim$\,3. The more recent study by \citet{bailey:2012} inferred $B_{\rm q}/B_{\rm d}=2.4$ for a stellar inclination similar to the one found in our study. These high $B_{\rm q}/B_{\rm d}$ values correspond to $>$\,80\% contribution of $\ell=2$ harmonic modes to the total magnetic energy. In addition, both our mean and maximum local magnetic field strengths are lower by about a factor of 2--3 than the values suggested by \citet{landstreet:1990} and \citet{bailey:2012}.

Unlike the previous magnetic field studies of \hr, which were limited to fitting longitudinal magnetic field measurements and magnetic line broadening, we considered the full Stokes $I$ and $V$ line profile information, took into account surface chemical inhomogeneities and derived consistent results from an independent modelling of three chemical elements. The resulting best-fitting MDI magnetic field maps successfully explain both the polarisation profiles themselves and their low-order moments, including the mean longitudinal magnetic field. Thus, our magnetic field geometry model passes a more rigorous observational test and therefore is more reliable. On the other hand, as shown above, the magnetic geometry proposed by \citet{bailey:2012} fails to reproduce the amplitude and detailed shapes of the Stokes $V$ profiles.

\begin{figure*}[!th]
\centering
\firrps{15cm}{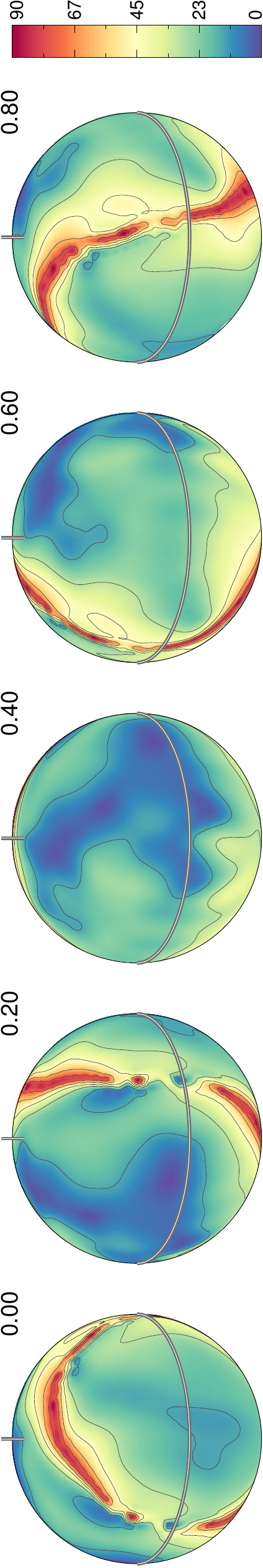}
\caption{Angle between the local field vector and the surface normal corresponding to the magnetic field maps reconstructed from the 
Fe LSD profiles (see Fig.~\ref{fig:fld_fe}). The contours over spherical maps are plotted with a step of 15\degr.}
\label{fig:fld_ang}
\end{figure*}

\begin{figure}[!th]
\centering
\fifps{8cm}{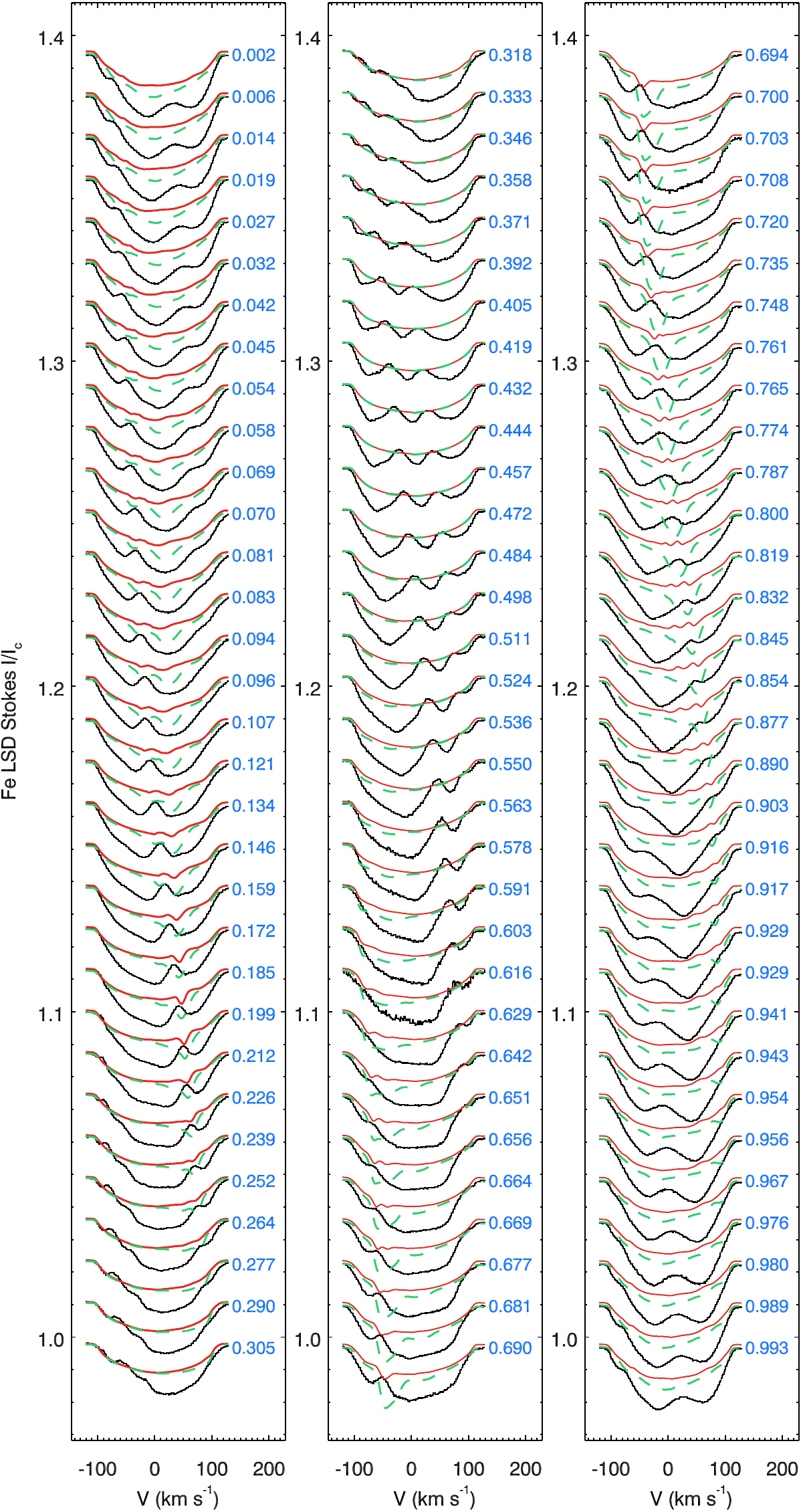}
\caption{Comparison of the observed Fe LSD Stokes $I$ profiles (black histogram) with calculations according to the diffusion theory predictions. The red solid lines show theoretical profiles for the iron distribution with $\log N_{\rm Fe}/N_{\rm tot}=-2.0$ in the regions with the local field inclination $\ge75\degr$ and $\log N_{\rm Fe}/N_{\rm tot}=-3.5$ elsewhere. The green dashed lines correspond to similar calculation with $\log N_{\rm Fe}/N_{\rm tot}=-2.0$ adopted for the surface regions with the field inclination $\ge45\degr$. 
Spectra corresponding to different rotation phases are offset vertically. Rotation phases are indicated to the right of each spectrum.}
\label{fig:prf_dif}
\end{figure}

The magnetic field analysis by \citet{bailey:2012} included an attempt to estimate the mean magnetic field modulus from the differential Zeeman broadening of two pairs of spectral lines with different effective Land\'e factors. This methodology, expected to be reliable for \vs\ of up to 50~\kms\ \citep{bailey:2014}, yielded mean field strength in the range of 11--23~kG for \hr, which is incompatible with our results. Examining synthetic and observed spectra of \hr\ in the vicinity of the four \ion{Cr}{ii} and \ion{Fe}{ii} lines used by \citet{bailey:2012}, we find that the profiles of all these absorption features are strongly blended by other lines and are significantly distorted by chemical abundance inhomogeneities. This makes it impossible to reliably isolate the \ion{Cr}{ii} and \ion{Fe}{ii} lines in question, let alone measure their widths with the precision of better than 0.5\% as implied by 2.5~kG field strength errors quoted by \citet{bailey:2012}. Therefore, it is likely that their mean field modulus measurements are entirely spurious.

Compared to the previous restricted parametric low-order multipolar models used by \citet{landstreet:1990} and \citet{bailey:2012}, our MDI code allows many more degrees of freedom corresponding to all possible poloidal and toroidal modes in a general spherical harmonic expansion. We also impose an additional regularising constraint forcing the global magnetic energy to a minimum, which dampens contributions of higher-order modes. Perhaps, in hindsight, it is not surprising that a model with a larger number of free parameters yields a weaker overall magnetic field with a lower quadrupolar contribution. Nevertheless, the fact that this model fits the observed polarisation profiles so successfully indicates that the quadrupole-dominated field structure previously proposed for \hr\ was merely a consequence of adopting a specific, highly restrictive low-order multipolar field parameterisation rather being linked to particular characteristics of the observational data. In this respect, our MDI analysis of \hr\ follows a series of polarisation profile studies of Ap/Bp stars which demonstrate frequent failures of the classical multipolar field models based on fitting the longitudinal field curves and other observables derived from Stokes profiles to match the line profile themselves \citep{bagnulo:2001,kochukhov:2004d,kochukhov:2011a}. It appears that predictions of such models are fairly robust only when dealing with simple, dipole-dominated field geometries but are becoming increasingly misleading whenever they attempt to incorporate large deviations from a dipolar field structure.

We may also compare the results of our detailed mapping with the chemical abundances derived by \citet{bailey:2012}. Because of the strongly different modelling assumptions, detailed comparisons are difficult to achieve. However, we note that in the case of the elements Mg, Si, Cr, and Fe the abundances derived by \citeauthor{bailey:2012} lie within the range of abundances in our maps, although generally with a smaller range of variation (consistent with the very low spatial resolution of \citeauthor{bailey:2012} surface mapping). In the cases of Ti and Nd, the abundances derived by \citeauthor{bailey:2012} are near the low abundance end of the range found here. Thus in general, chemical abundances derived from the simple model seem to be fairly indicative of our much more accurate values, and in any case do not lead to spuriously large overabundances.

To summarise, the main outcome of our MDI analysis of \hr\ is that the dominant contribution to its field topology is provided by a distorted dipole rather than an axisymmetric (linear) quadrupole. Polarisation profile modelling of HD\,37776, also believed to have a very strong quadrupolar field, yielded similar results, although in that case the best-fitting magnetic map was comprised of smaller-scale magnetic spots without an underlying dipole-like structure \citep{kochukhov:2011a}. Considering that \hr\ and HD\,37776 were the best candidates for a quadrupole-dominated field topology known among Ap/Bp stars, there seems to be no remaining evidence that such hypothetical magnetic field geometries occur in real stars. Interestingly, this conclusion is in line with the results of three-dimensional numerical simulations of the fossil magnetic field evolution in stably stratified stellar interiors. The studies by \citet{braithwaite:2006} and \citet{braithwaite:2008} have demonstrated that initially random magnetic field quickly develops into either a dipole-like topology or a complex, non-axisymmetric configuration, depending on the initial radial distribution of the field energy. The surface magnetic field structures of the dipole-like models calculated by Braithwaite (private communication) can, in fact, be quite distorted showing various dipole offsets, toroidal contributions and smaller-scale magnetic features not unlike the pair of spots at the negative magnetic pole of \hr\ or those found in the Stokes $IQUV$ inversions of $\alpha^2$~CVn \citep{kochukhov:2010,silvester:2014}. On the other hand, none of these theoretical MHD models yields a surface field configuration reminiscent of a linear quadrupole.

\subsection{Comparison with diffusion theory predictions}

Atomic diffusion in the presence of a magnetic field is believed to the main mechanism responsible for non-solar chemical abundance patterns and chemical inhomogeneities in the surface layers of Ap/Bp stars \citep{michaud:2015}. Equilibrium diffusion models, such as those calculated by \citet{leblanc:2009} and \citet{alecian:2010}, predict significant vertical element stratification in the atmospheric line-forming region. These vertical stratification profiles presumably change across the stellar surface according to the local magnetic field geometry, giving rise to the rotational line profile modulation interpreted by DI codes as chemical spots. As summarised by \citet{ryabchikova:2008a} and \citet{ryabchikova:2011}, stratification profiles predicted by diffusion calculations agree qualitatively and, in some cases, quantitatively with the findings of observational studies, especially those targeting cool Ap stars which exhibit the most pronounced vertical abundance gradients \citep[e.g.][]{kochukhov:2006b,ryabchikova:2008,shulyak:2009,nesvacil:2013}. 

On the other hand, no satisfactory agreement can be found between the observed DI maps and theoretically predicted horizontal abundance structures. It appears that the current equilibrium diffusion calculations predict no variation of chemical stratification over the stellar surface except in narrow bands coinciding with horizontal field regions. For example, according to \citet{alecian:2010}, the iron vertical stratification profiles calculated for $T_{\rm eff}=12000$~K, $\log g=4.0$ atmospheric model are independent of the field orientation until the field vector is inclined by more than 75--80\degr\ with respect to the local surface normal. Almost all other chemical elements, including Si and Mg studied here, are predicted to follow the exact same pattern, which would be interpreted by Doppler inversions as narrow overabundance rings located at the magnetic equator. More sophisticated time-dependent diffusion calculations \citep{alecian:2011,stift:2016} have yet to be published in a format enabling a direct comparison with MDI maps. However, despite some changes in the vertical element distributions, there are no indications that these time-dependent calculations would lead to any other horizontal pattern besides narrow overabundance rings.

Taking advantage of the detailed information about the magnetic field geometry of \hr\ and several chemical abundance maps reconstructed self-consistently with the magnetic field, we attempted to carry out a quantitative comparison of our observational data with the theoretical element distribution models initially presented by \citet{alecian:2010} and updated by \citet{alecian:2015}. It should be emphasised that \hr\ is a particularly suitable target for such a comparison due to its high projected rotational velocity. For stars with low \vs\ one can plausibly argue that lateral variation of the vertical stratification profile might lead to changes of the local line shapes that interfere with the usual DI inversion based on the assumption of a single value of abundance for each surface point. However, for stars with \vs\ as high as $\sim$100~\kms\ details of the local line profiles are irrelevant \citep[e.g.][]{unruh:1995}, meaning that the standard DI inversion should provide a reliable estimate of the mean local element abundance.

Given the high quality of the Fe and Si LSD Stokes $I$ profiles obtained for \hr, we decided to assess diffusion theory predictions for these two elements. However, despite the presence of bi-dimensional Si maps in the publication by \citet{alecian:2010}, no meaningful predictions can be made for this element according to \citet{alecian:2015} because Si is unsupported by the radiative pressure everywhere on the stellar surface. This results in a sub-solar abundance of this element, contrary to the well-established large overabundance of Si in \hr\ and many other Si-rich Bp stars. Limiting our analysis to Fe alone, we calculated the local field orientation according to our Fe MDI results. The corresponding absolute field inclination map, shown in Fig.~\ref{fig:fld_ang}, agrees very well with the field inclination distributions inferred from the Si and Cr LSD profile inversions. This map shows a slightly warped ring of high field inclination wrapped around the star. The inferred location of the horizontal field zones is very robust since it is mostly determined by the radial field component which is well-constrained for a dipole-like field even in the absence of the Stokes $Q$ and $U$ data.

Considering the Fe field inclination map, we adopted $\log N_{\rm Fe}/N_{\rm tot}=-2.0$ for the regions where the field inclination exceeded 75\degr\ and $\log N_{\rm Fe}/N_{\rm tot}=-3.5$ elsewhere. These abundances were adopted to approximately represent the inferred maximum and mean Fe concentration, respectively. The exact abundance values are less important for the following qualitative discussion of line profile variability compared to the position and size of element overabundance regions.

Unsurprisingly, given the tiny surface fraction (about 3\%) occupied by field of high inclination and therefore high Fe abundance, this element distribution model yields a very weak, small-scale Stokes $I$ line profile variation (red solid line in Fig.~\ref{fig:prf_dif}), which is entirely unlike the broad, high-amplitude distortions travelling across the observed Fe line profiles. Trying to increase the area of Fe overabundance regions, we generated another map with $\log N_{\rm Fe}/N_{\rm tot}=-2.0$ adopted for the zones with field inclination exceeding 45\degr. The resulting profile variation, shown by the dashed line in Fig.~\ref{fig:prf_dif}, is more discernible but is still vastly different compared to observations. In fact, there is an anti-correlation between dips in the model profiles and bumps in observations, indicating that Fe is underabundant instead of overabundant in the regions characterised by a high field inclination.

According to the diffusion calculations at $T_{\rm eff}=10000$~K \citep{alecian:2015}, the Cr abundance distribution is expected to be very similar to the one for Fe, i.e. it is dominated by a narrow overabundance ring coinciding with the magnetic equator. Given the similarity of the Cr and Fe line profile variation in \hr, this hypothetical chromium distribution would evidently lead to the same disagreement with observations as discussed above for Fe. On the other hand, earlier calculations by \citet{alecian:2010} seem to indicate almost no horizontal Cr abundance variation at $T_{\rm eff}=12000$~K, in conflict with the high-contrast Cr distribution map inferred in our study and with conspicuous Cr line variability often observed in Bp stars in this temperature range.

It is beyond the scope of this paper to address the reasons for this apparent failure of the diffusion theory to match the observed surface abundance structures. Several physical processes that can potentially alter predictions of the current diffusion models have been enumerated by \citet{alecian:2015}. Among various effects currently overlooked by these calculations, the anisotropic mass-loss, already known to be important in the context of atmospheric diffusion \citep{babel:1992}, is particularly relevant. Perhaps, a more successful diffusion models can be developed by imposing the outer atmospheric boundary condition according to the magnetically confined \citep{babel:1997,townsend:2005}, chemically fractionated \citep{krticka:2014} stellar wind treatment. In addition, the current-driven diffusion effects \citep{urpin:2016} and the global hydrodynamical instabilities likely responsible for the low-contrast dynamic spots on HgMn stars \citep{kochukhov:2007b,korhonen:2013} should be seriously looked into. In any case, our assessment shows that the current theoretical diffusion modelling of the horizontal chemical abundance structures is far too simplistic and needs to be significantly improved before the output of these calculations could be meaningfully compared to the empirical DI maps of Ap/Bp stars.

\section{Conclusions}
\label{conc}

In this paper we presented the first DI and MDI analysis of \hr. This is a young, rapidly rotating, magnetic B-type chemically peculiar star with a well-known age due to its cluster membership. \hr\ is famous for its unusual double-wave mean longitudinal magnetic field curve, which was repeatedly interpreted as a signature of an exceptionally strong, quadrupole-dominated magnetic field topology.

Using a set of nearly one hundred Stokes $I$ and 52 Stokes $V$ high resolution observations acquired for \hr\ with the HARPSpol instrument at the ESO 3.6-m telescope, we found that the previously proposed quadrupolar field topology models are inconsistent with the observed polarisation profiles of metal lines. Interpretation of the LSD Stokes $V$ profiles with an MDI code points to an asymmetric, bipolar field geometry with a large, weak positive field region and a smaller, strong negative field zone comprised of two distinct magnetic spots with a somewhat different field orientation. An independent MDI analysis of Si, Cr, and Fe lines yields consistent magnetic field maps, indicating the mean local field strength of 4.0--4.4~kG and the maximum local field of about 12~kG at the negative magnetic pole. The field geometry is dominated by the dipolar component, which accounts for 65--71\% of the total magnetic field energy. No evidence of a significant toroidal field is found in our magnetic mapping results.

The distorted, non-axisymmetric dipolar, rather than axisymmetric quadrupolar, magnetic field structure of \hr\ qualitatively agrees with numerical simulations of fossil magnetic field evolution in stably stratified stellar interiors. Our results obtained for \hr\ together with other recent MDI studies suggest that linear quadrupolar fields, often employed for describing Ap/Bp star magnetic field topologies, do not exist in real stars. Instead, the surface fossil field structure appears either as a (sometimes strongly distorted) dipole in the majority of stars or as a far more complex superposition of magnetic spots in few objects.

We have derived horizontal distributions of six chemical elements, including Si, Cr, Fe, Mg, Ti, and Nd, using LSD Stokes profiles and intensity profiles of individual spectral lines. These abundance maps are characterised by high-contrast, large-scale overabundance features, which show no simple correlations with the local magnetic field strength and orientation. We showed that the observed Fe and Cr profile variation contradicts expectations of the current theoretical diffusion models, which predict narrow overabundance zones coinciding with the magnetic equator for the iron-peak and almost all other chemical elements. We suggest that this controversy should be resolved by making theoretical models more realistic and, in particular, incorporating an anisotropic mass loss in diffusion modelling.

\begin{acknowledgements}
O.K. acknowledges financial support from the Knut and Alice Wallenberg Foundation, the Swedish Research Council, and the Swedish National Space Board.
G.A.W. is supported by a Discovery Grant from the Natural Science and Engineering Research Council (NSERC) of Canada.
\end{acknowledgements}


\end{document}